\documentclass[useAMS,usenatbib]{mn2e}
\usepackage{amssymb}
\usepackage{amsmath}
\usepackage{multirow}
\usepackage{graphicx}
\usepackage{tikz}

\def\lya{Ly$\alpha$}
\def\lyc{LyC}
\def\filterb{\textit{B}}
\def\filterv{\textit{V}}
\def\filterr{\textit{R}}
\def\filteri{\textit{i}$'$}
\def\filterz{\textit{z}$'$}
\def\filterlyc{\textit{NB359}}
\def\filterlya{\textit{NB497}}
\def\filterbv{\textit{BV}}

\def\ew{$EW($\lya$)$}

\def\papi{{\sevensize Paper}\,{\sevensize I}}
\def\sex{{\sevensize SExtractor}}

\title[LyC AGN in the SSA22 field]
  {Lyman continuum leaking AGN in the SSA22 field\thanks{Based on data collected at Subaru Telescope, which is operated by the National Astronomical Observatory of Japan}}
\author[Micheva et al.]
  {Genoveva Micheva,$^1$
  Ikuru Iwata,$^{1,2}$ and
  Akio K. Inoue$^3$\\
  $^1$Subaru Telescope, National Astronomical Observatory of Japan,
  650 North A'ohoku Place, Hilo, HI 96720, USA\\
  $^2$Graduate University for Advanced Studies (SOKENDAI), Osawa 2-21-1, Mitaka, Tokyo 181-8588, Japan \\
  $^3$College of General Education, Osaka Sangyo University,
  3-1-1, Nakagaito, Daito, 574-8530 Osaka, Japan.\\
     }
\date{Released 2015 Xxxxx XX}

\pagerange{\pageref{firstpage}--\pageref{LastPage}} \pubyear{2015}

\def\LaTeX{L\kern-.36em\raise.3ex\hbox{a}\kern-.15em
    T\kern-.1667em\lower.7ex\hbox{E}\kern-.125emX}

\begin{document}

\label{firstpage}

\maketitle

\begin{abstract}
Subaru/SuprimeCam narrowband photometry of the SSA22 field reveals the presence of four Lyman continuum (\lyc) candidates among a sample of $14$ AGN. Two show offsets and likely have stellar \lyc~in nature or are foreground contaminants. The remaining two \lyc~candidates are Type I AGN. We argue that the average \lyc~escape fraction of high redshift low luminosity AGN is not likely to be unity, as often assumed in the literature. From direct measurement we obtain the average \lyc-to-UV flux density ratio and ionizing emissivity for a number of AGN classes and find it at least a factor of two lower than values obtained assuming $f_{esc}=1$. Comparing to recent Ly$\alpha$ forest measurements, AGNs at redshift $z\sim3$ make up at most $\sim12\%$ and as little as $\sim5\%$ of the total ionizing budget. Our results suggest that AGNs are unlikely to dominate the ionization budget of the Universe at high redshifts. 
\end{abstract}


\section{Introduction}
\noindent The intergalactic medium (IGM) is ionized and kept ionized by quasars (QSOs)/AGN and star forming (SF) galaxies. Identifying the primary population of ionizing radiation (a.k.a. Lyman Continuum, \lyc) at any redshift has long been the focus of intensive studies and the situation is far from settled. The neutral hydrogen in the IGM seems to dramatically increase beyond redshift $z>6$~\citep[e.g.][]{2001AJ....122.2833F,2001AJ....122.2850B,2002AJ....123.2151P}, while the IGM appears to be highly ionized at $z\lesssim6$. Some research suggests that QSOs dominate the ionization budget at redshifts $z\lesssim3$~\citep[e.g.][]{1996ApJ...461...20H,2014MNRAS.438.2097F,2015MNRAS.451L..30K,2015arXiv150901101G}, others suggest that SF galaxies need to significantly contribute even at these redshifts~\citep[e.g.][]{2009ApJ...703.1416F,2011ApJ...728L..26G}, or that the contribution of both QSOs and SF galaxies is comparable~\citep[e.g.][$z<2$]{2015ApJ...811....3S}. Observational evidence suggests that SF galaxies do not produce enough ionizing radiation at redshifts $z\leq3$~\citep[e.g.][]{2015arXiv150901101G} to maintain the ionization levels, but that their ionizing emissivity substantially increases with increased redshift~\citep[e.g.][]{2006MNRAS.371L...1I,2013MNRAS.436.1023B,2015MNRAS.451.2030D} and that they alone ionize the Universe at high redshifts~\citep[e.g.][]{2010ApJ...710.1239R,2012ApJ...759L..38A,2015ApJ...811..140B,2015MNRAS.451.2030D}. The exact ``turnover'' redshift at which SF galaxies (including high-z low luminosity dwarfs) take over the production of ionizing photons is unclear, since their contribution seems to not be high enough to dominate the ionizing emissivity at $z=3$~\citep[e.g.][]{2015arXiv150901101G} or even $z>4$~\citep[e.g.][]{2009ApJ...692.1476C}. On the other hand the luminosity function of bright QSOs suggested until recently that they are not numerous enough to dominate at $z>~6$~\citep[e.g.][]{2005MNRAS.356..596M,2012MNRAS.425.1413F,2014MNRAS.438.2097F}. The situation may be different for faint AGN, however, for which recent research suggests a less steep decline of their number density at high redshifts. Faint AGN may in fact be numerous enough to be the main contributors to the ionizing emissivity at redshifts $z>4$~\citep{2012ApJ...755..124G,2015A&A...578A..83G,2015ApJ...813L...8M}.\\
\setcounter{table}{0}
\begin{table*}
\begin{minipage}{184mm}
\centering
\caption{Summary of confirmed AGN in our sample. Where available the Chandra catalog number (column L09) is indicated with the arcsec distance for coordinate cross-matching (column dist). }\protect\label{tab:var}
\begin{tabular}{ l l l c l r  c c r r r}
\hline
ID & RA & DEC & z & z Ref & L09 &  dist & Type &  S   & P[\%] & Alt. name\\
\hline
AGN1  &  22:17:5.4   &  0:15:14.0               &   3.801  & $\star$          &  12   & 0.5    &1  & 1.8  &   32.8 &-\\
AGN2  &  22:17:5.8   &  0:22:25.2               &   3.083  & $\star$          &  15   & 1.0    &2  & 0.7  &   75.6 &-\\
AGN3${}^{\dagger}$  &  22:17:6.7   &  0:26:41.1   &   3.140  & $\star$          &  -    &  -     &1  & 1.2  &   34.8 & -\\
AGN4${}^{\dagger}$  &  22:17:9.6   &  0:18:0.7    &   3.106  & Y12,L09,$\star$  &  20   &  0.8   &2  & 1.0  &   50.6 & LAE J221709.6+001801 \\
AGN5${}^{\dagger}$  &  22:17:12.7   &  0:28:55.6  &   3.110  & $\star$          &  -    &  -     &1  & 3.8  &    0.0 &  [IKI2011] i\\
AGN6  & 22:17:16.2 & 0:17:44.47                 &   3.098  & S15              &  43   &  1.7   &2  & 1.0  &   75.9 & CXOSSA22 J221716.1+001745 \\          
AGN7  &  22:17:20.2   &  0:20:19.1              &   3.108  & S15              &  57   &  0.2   &1  & 7.4  &    0.0 & LAE J221720.2+002019 \\
AGN8  &  22:17:22.3   &  0:16:40.1              &   3.353  & S15,VVDS         &  68   &  0.1   &1  & 3.3  &    0.1 & SSA 22a D013 \\
AGN9  &  22:17:25.4   &  0:17:16.8              &   3.105  & N13              &  80   & 0.2    &2   &   ?  &   ?    & [NSS2011] LAE 017 \\
AGN10  &  22:17:35.8   &  0:15:59.0              &   3.094  & S15             &  139  & 0.2    &2  & 0.4  &   85.0 & [GMS2005] LAB 14 \\
AGN11${}^{\dagger}$  &  22:17:36.5   &  0:16:22.6 &   3.084  & S15,VVDS         &  140  & 0.0    &1  & 2.5  &   11.6 & SSA 22a D012 \\
AGN12  &  22:17:39.1   &  0:13:29.8             &   3.091  & W05              &  153  & 0.9    &2  & 1.1  &   61.7 & [WYH2009] LAB 02 b\\
AGN13  &  22:17:51.3   &  0:20:38.4             &   3.460  & S15              &  239  & 0.1    &1  & 1.1  &   64.3 & -\\
AGN14  &  22:17:59.2   &  0:15:29.5             &   3.094  & S15              &  268  & 0.4    &2  & 1.3  &   65.7 & -\\
\hline
\end{tabular}
\end{minipage}
\medskip 
$\dagger$ -  object has been detected in the \filterlyc~(LyC) band; $\star$ - redshift and type from spectrum in Figure~\ref{fig:spectra} in the appendix; $S$ - variability significance defined in~\citet{2007ApJ...665..225K}; $P$ - median probability from $10000$ MC simulations for a random match to the individual light curve profiles; Type 1: broad emission lines in spectrum; Type 2: narrow emission lines;  IKI2011 -~\citet{2011MNRAS.411.2336I}; GMS2005 -~\citet{2005MNRAS.363.1398G}; NSS2011 -~\citet{2011ApJ...736...18N}; WYH2009 -~\citet{2009ApJ...692.1561W}; Y12 -~\citet{2012AJ....143...79Y}; S15 -~\citet{2015MNRAS.450.2615S}; L09 -~\citet{2009MNRAS.400..299L}; N13 -~\citet{2013ApJ...765...47N}; W05~\citet{2005Natur.436..227W}; VVDS - VIMOS VLT Deep Survey.
\end{table*}

\noindent Much of the literature on \lyc~from AGN relies on power law parameterizations of observations of quasar spectral energy distributions (SEDs). The continuum slopes can vary widely~\citep[e.g.][]{1997ApJ...475..469Z,2001AJ....122..549V,2002ApJ...565..773T,2004ApJ...615..135S,2015A&A...578A..83G,2015MNRAS.449.4204L,2016arXiv160309351C}, and often it is assumed that the average escape fraction of \lyc~radiation from AGN/QSOs is $\left<f_{esc}\right> =1$~\citep[e.g.][]{2015A&A...578A..83G,2015ApJ...813L...8M}. In this paper we take a look at a sample of AGN in the SSA22 field and analyze the detected \lyc~emission from Subaru/SuprimeCam narrowband data. We present an average ionizing emissivity from direct observations, and the \lyc~escape fraction relative to an assumed intrinsic spectrum. \\

\noindent Throughout this paper we use the AB magnitude system and adopt a flat $\Lambda$CDM cosmology with $H_0 = 70$ km s${}^{-1}$ Mpc${}^{-1}$, $\Omega_{M}= 0.3$, $\Omega_{\Lambda}= 0.7$.

\section{Sample selection}
We assembled a catalog of AGN from the SSA22 field by first searching among our base sample of $308$ galaxies with $z>3.06$ from~\citet[][hereafter \papi]{2015arXiv150903996M}. With this redshift cutoff the \filterlyc~filter samples \lyc~radiation. Initially we identified $7$ objects with broad emission line AGN spectral signatures, i.e. AGN of Type I. \citet{2008A&A...492..637G} published an updated catalog of the VIMOS VLT Deep Survey (VVDS) of broad-line (type-1) AGN with a total of $298$ objects, $41$ of which have redshifts $z\ge3.06$. The only matches with our catalog are two objects already classified as broad emission line AGN from our spectra, AGN8 (SSA22a-D13) and AGN11 (SSA22a-D12). Cross-matching with the Chandra Deep Protocluster Survey catalog of Xray sources in the SSA22 field~\citep{2009MNRAS.400..299L} we found $12$ Xray AGN among our sample. The coordinate matching tolerance for all but one object was $\le1.0\arcsec$ and eight of these sources were matched to better than $0.5\arcsec$. Among these Xray AGN five were already known to us as Type I AGN from their spectra while the other six were previously classified as LAEs or LBGs in our sample.\\

\noindent Recently, \citet{2015MNRAS.450.2615S} performed a spectroscopic survey of the SSA22 field with Keck and VLT, looking to identify new protocluster members and to complement literature studies with robust redshifts. They provide spectra of $7$ Xray AGN which contain one new Xray AGN match with our catalog. This is AGN6, with the largest coordinate matching tolerance of $1.7$\arcsec, added to our sample because the Xray emission is extended and occupies a region which contains the optical coordinates of the object in our sample (Saez, private communication). The total number of unique Type I or Xray AGN in our sample is thus $14$, summarized in Table~\ref{tab:var} and with complementary to the literature spectra in Figure~\ref{fig:spectra} in the appendix. Table~\ref{tab:var} also shows our complementary classfication of the AGN in our sample, using a $\geq1000$ km s${}^{-1}$ velocity width cutoff for Type I AGN, and classifying all non-Type I objects as being Type II since the quality of the spectra does not allow for a finer classification.\\

\noindent Among all $14$ AGN four were also detected in the \filterlyc~narrowband data within $\leq0.8$\arcsec of the \filterr~band detection, so these consistitute our initial sample of \lyc~AGN. Three of them are Type I (AGN3, AGN5, AGN11), and one is Type II (AGN4). The spectrum of AGN4 in the appendix shows a rest-frame velocity width of both the \lya~and the {\sevensize CIV} lines of $\sim670$ km s${}^{-1}$, and is thus clearly not Type I.~\citet{2009ApJ...692.1476C} observe that only Xray sources with broad emission lines in their spectra have detectable ionizing flux. The ionizing radiation from AGN4 could be stellar in nature or this could be a foreground contaminant.\\

\begin{figure}
\centering\tiny
\includegraphics[height=73mm]{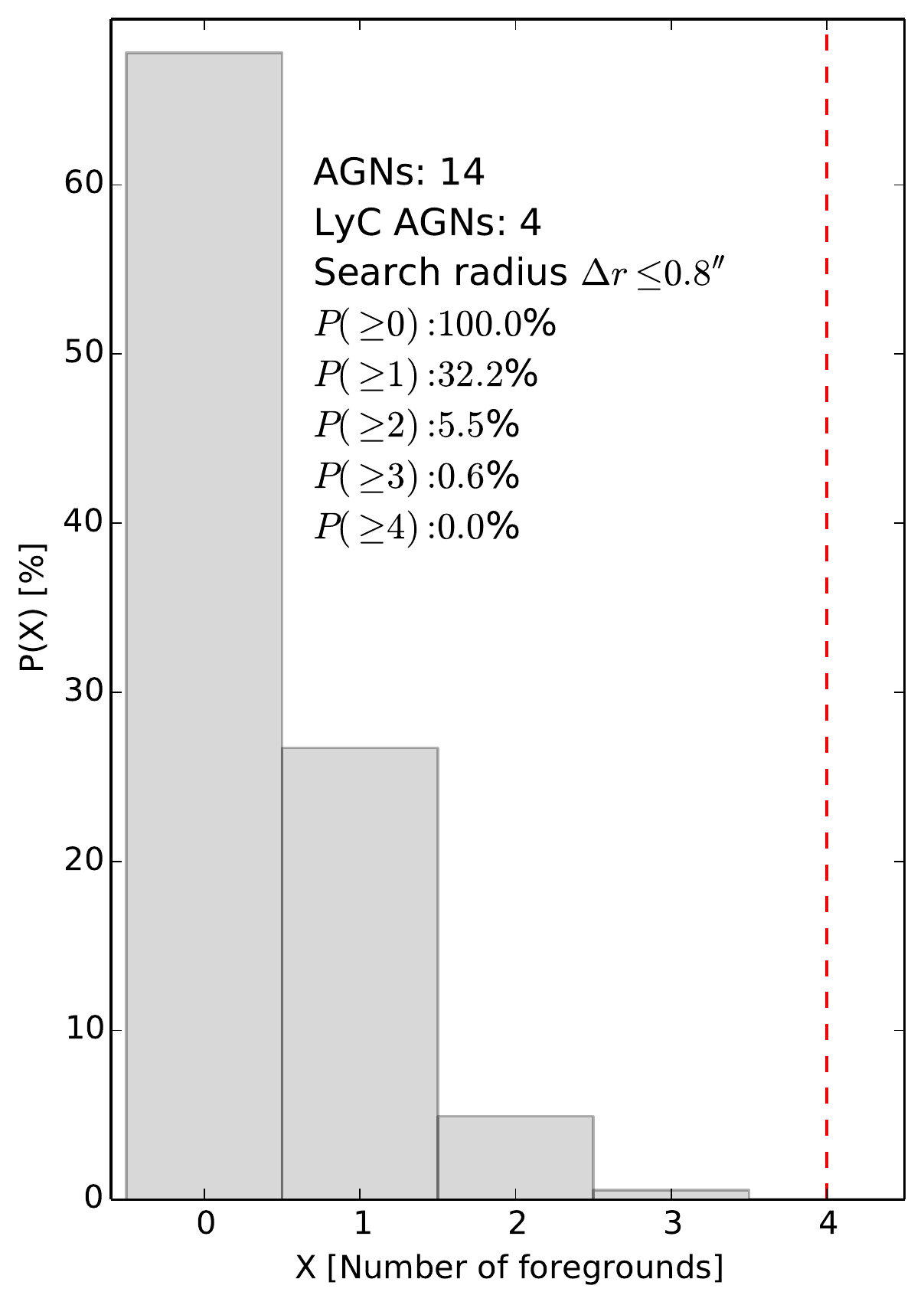}
\caption{Probability mass function of the expected number of foreground contaminants from an MC simulation with 100000 runs. Inset text shows cumulative probabilities to obtain e.g. at least $X$ or more contaminants. The most likely number of foreground contaminants is zero, with $P(X=0)=68\%$. }\protect\label{fig:contaminants}
\end{figure}

\noindent In Figure~\ref{fig:contaminants} we show the probability mass function (PMF) of the expected number of foreground contaminants using the same Monte Carlo (MC) method from \papi~which utilizes the spatial distribution of all detections in our widefield \filterlyc~image. The search radius used in this MC simulation is $r=0.8$\arcsec, the largest offset we measure in the \lyc~AGN candidate sample. The most likely number of contaminants is zero, with a probability of full contamination of $P(\ge4)=0.03\%$. \\

\section{Variability}

\noindent Our multiwavelength data were not taken all at the same epoch. To reliably interpret the photometry of the AGNs, and specifically their ionizing to non-ionizing flux density ratios from mixed epochs, we need to determine if they are significantly variable. Additionally, \citet{2015A&A...576A.116V} suggest that low luminosity AGN hidden in star forming galaxies could be responsible for the detected \lyc~emission from these galaxies. We want to check if, given the quality of our data, we can detect variability among confirmed, faint AGN. This would complement \papi, where using the same data we tested candidates of \lyc-leaking star-forming galaxies at $z\sim3.1$, and saw no variability among them.\\
\begin{figure}
\centering\tiny
\includegraphics[width=85mm]{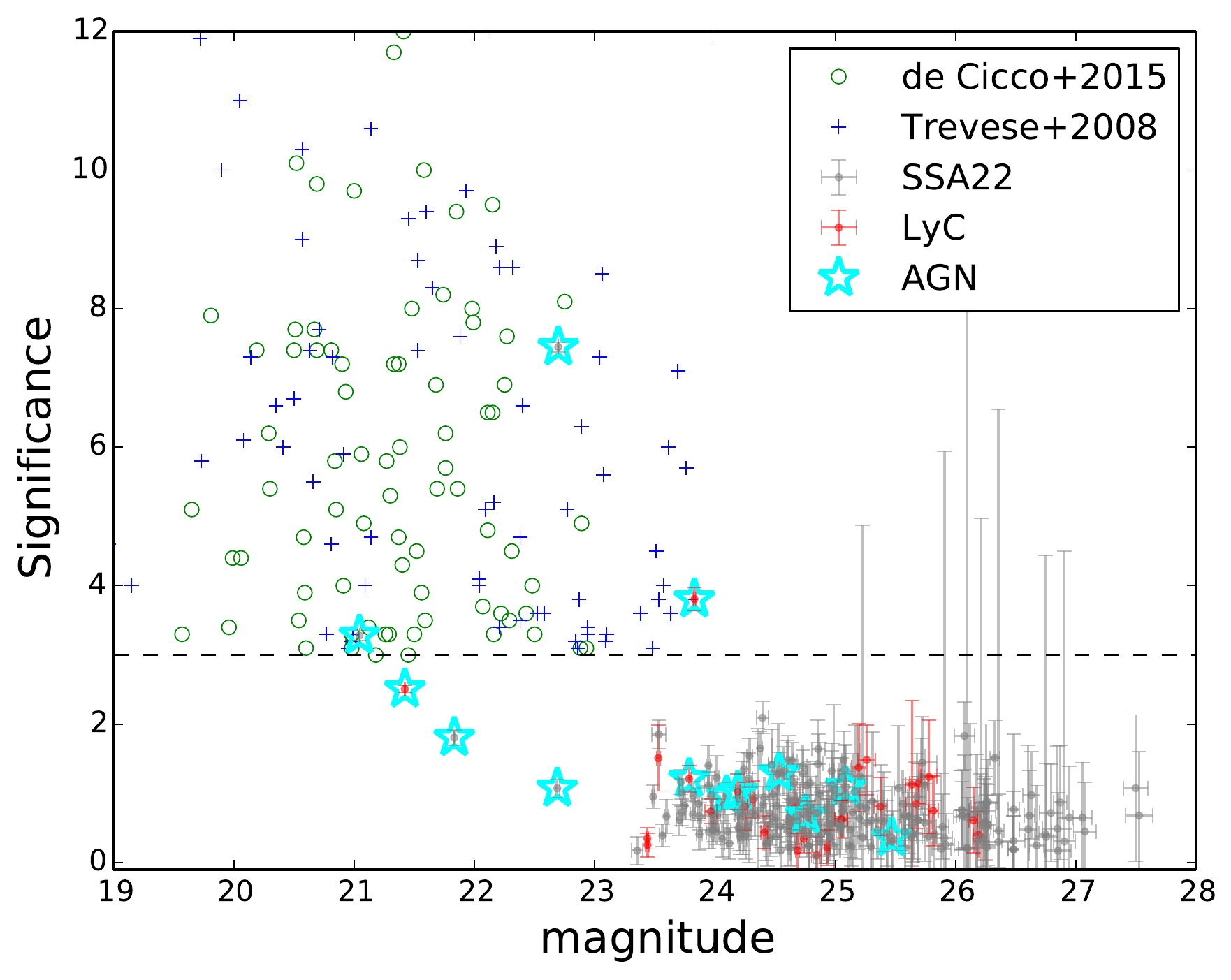}
\caption{Variability significance versus rest-frame UV magnitude (\filterr~band) for the SSA22 sample. The Xray or spectroscopically confirmed AGN in our sample are indicated with cyan stars. The dashed line indicates the variable/non-variable separation adopting $S=\sigma/ err_{\sigma}\ge3$ as variability criterion. Variable AGNs from the VST COSMOS field~\citep{2015A&A...574A.112D} and Chandra Deep Field South~\citep{2008A&A...488...73T} are added for comparison.}\protect\label{fig:KS07}
\end{figure}

\noindent Xray and UV variability among AGN has been well documented in the literature in AGNs with high and intermediate-mass black holes (BH), both on short (seconds/days) and long (months/years) time scales~\citep[e.g.][]{1997ARA&A..35..445U,2003ApJ...599..933N,2003AJ....126.1217D,2005AJ....129..615D,2005ApJ...631..741G,2005A&A...443..451F,2005MNRAS.359.1469M,2007ApJ...665..225K,2008ApJ...686..892T,2009MNRAS.394..443M,2012A&A...542A..83P,2015MNRAS.447.2112L}. The dominant variable component seems to be due to variability in the flux of the continuum~\citep{2015MNRAS.447...72P}, and although variability decreases towards longer wavelengths~\citep[e.g.][]{2002ApJ...564..624T,2003AJ....126.1217D}, there seems to be no wavelength dependence on the variability timescale~\citep{2005A&A...443..451F}. Therefore, one can for example use observed UV, optical or IR long-term variability  to successfully identify potential AGN candidates~\citep[e.g.][]{2007ApJ...665..225K,2008A&A...488...73T, 2008ApJ...676..163M, 2011ApJ...731...97S,2015A&A...574A.112D,2015MNRAS.446.3199G}. The variation and its significance are often determined by the average magnitude of all epochs, the rms around the average and an estimate of the uncertainty, from for example formal photometric errors or based on empirical estimates of the magnitude spread of non-variable objects.\\

\noindent We investigated the variability of our data, specifically we looked for variability among the \lyc~candidates in \papi~and the known AGN in our sample. We have reduced Subaru/Suprime-Cam \filterb~band data of the SSA22 field taken in the years $2002,~2003,~2007,~2008$, resulting in one final frame per year. Using the NASA Extragalactic Database (NED) catalog $101$ stars were identified in each frame. The year $2008$ was taken as reference, the median offset in \sex~photometry between the reference and the rest of the years was calculated, and applied to the corresponding frames. The median offsets between the stellar magnitude measurements were $\delta(2008-2002)=0.27^{\textrm{m}}$, $\delta(2008-2003)=0.53^{\textrm{m}}$, $\delta(2008-2007)=0.31^{\textrm{m}}$. We also measured the median $FWHM$ from the stars in each frame, finding a maximum $FWHM=1.1\arcsec$ for the year $2007$.\\
\begin{figure}
\centering\tiny
\includegraphics[width=85mm]{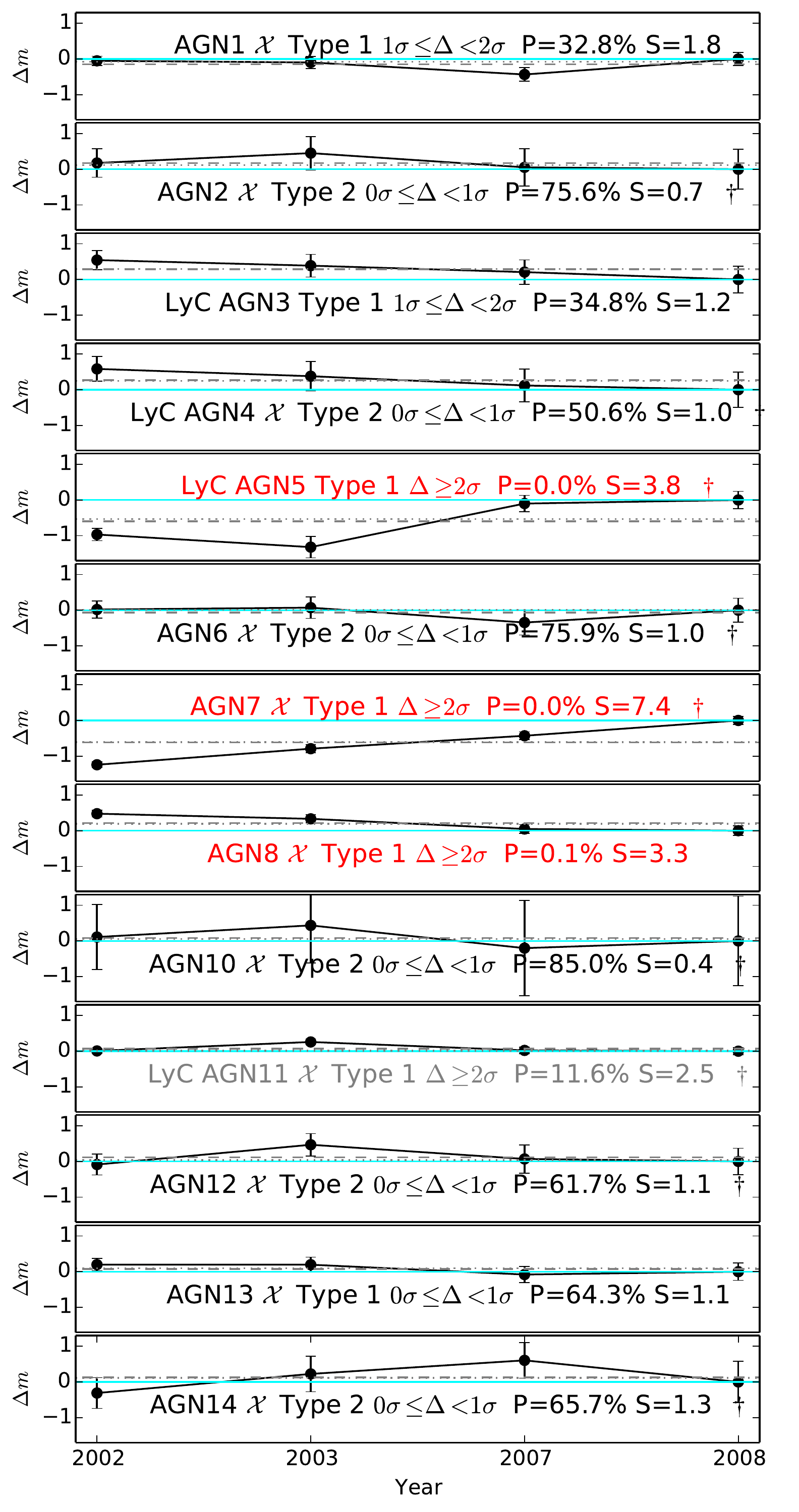}
\caption{Light curves for all AGN in the sample. Most AGN have no detectable variability (black text), one is marginally variable (gray text), and three are significantly variable (red text). The source ID and classification ($\mathcal{X}$= Xray source, Type 1 or 2) are indicated for each object, as well as the variability ``strength'' $\Delta$, the probability of random match to the current profile (P), and the significance (S) from the formal significance test. Protocluster association is indicated by $\dagger$. Solid cyan line is at the reference year $2008$ of $\Delta m=0$. The dotted (dashed) line is the median (average) of the 4 points.}\protect\label{fig:varconfirmed}
\end{figure}

\noindent After the zeropoints and PSFs between the $4$ years were equalized we performed \sex~photometry ({\sevensize MAGAUTO}) on our sample of $308$ LAEs, LBGs, and spectrally-confirmed AGN in \papi, all with spectroscopically confirmed redshifts $z\ge3.06$. There are several factors to keep in mind in our analysis. Our base sample is relatively small, especially compared to e.g. those of~\citet[][]{2008A&A...488...73T} and~\citet[][]{2015A&A...574A.112D} who have $\sim7000$, respectively $\sim18000$ objects. Our individual exposures are not particularly deep, in fact we only detect $225$ out of $308$ objects in at least three of the four individual epochs, and $<200$ in all four epochs. The bulk of our sample is several magnitudes fainter than the limits in~\citet[][]{2008A&A...488...73T} and~\citet[][]{2015A&A...574A.112D}, and has significant photometric uncertainties which should be taken into account. We therefore use the significance definition of~\citet{2007ApJ...665..225K}, $S=\sigma/err_{\sigma}$, with $err_{\sigma}=\sqrt{ \sum{ \sigma_{m}}^2 / N }$, and $\sigma_{m}$ as the formal photometric error for a magnitude measurement at a given epoch for $N$ epochs.\\

\begin{figure}
\centering\tiny
\includegraphics[height=40mm]{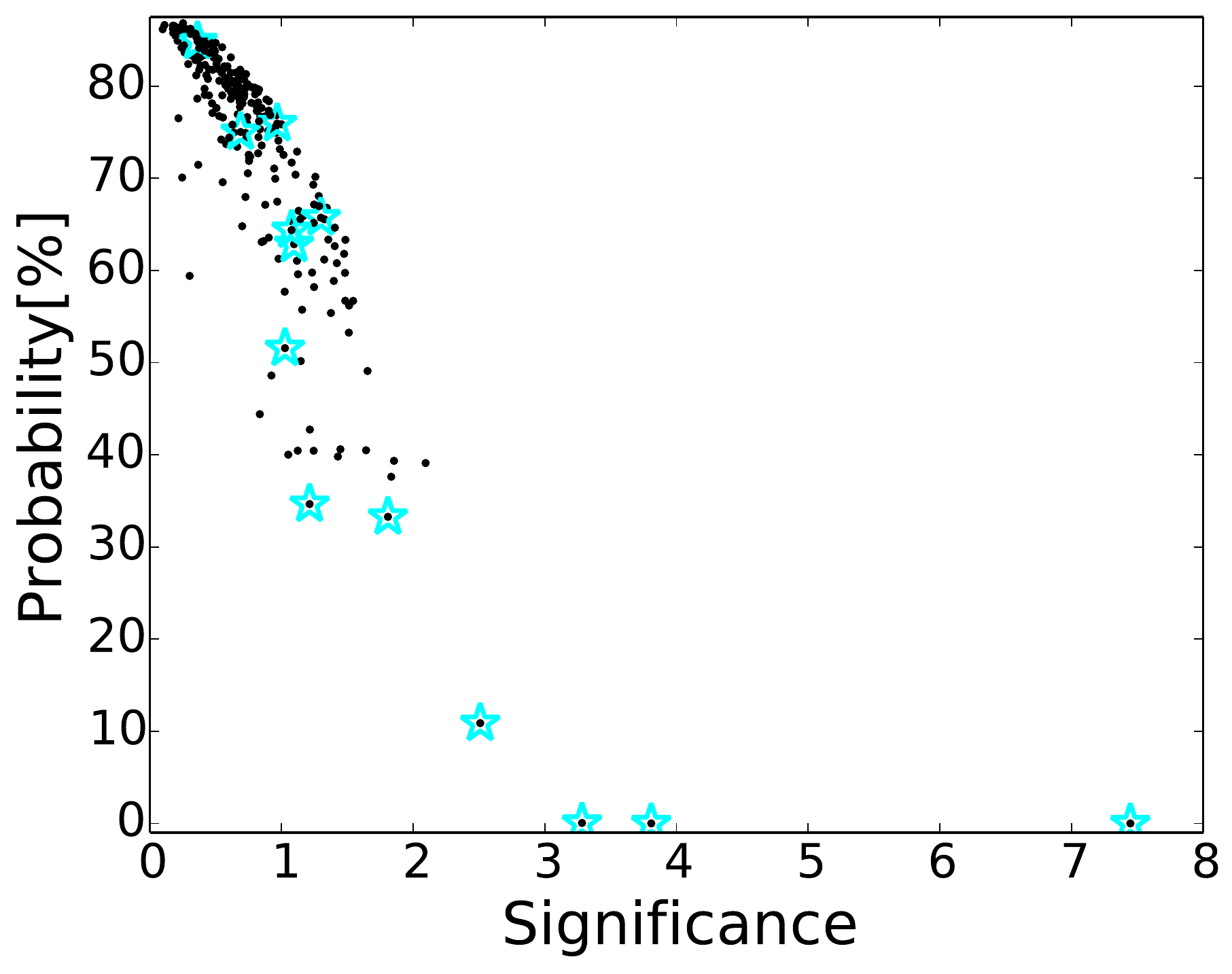}
\caption{Random-match probability P versus variability significance S. Confirmed AGN (cyan stars) are indicated for convenience. }\protect\label{fig:pvss}
\end{figure}

\noindent We note that many studies use small apertures of fixed radius ($\sim1\arcsec$) in an attempt to isolate the nuclei and reduce the diluting effect the host galaxy brightness has on the variability of the nucleus, while we use {\sevensize MAGAUTO}. Our previous analysis (\papi) suggested that many sources in our sample are extended and/or have several substructures. Since we do not in advance know which substructure may contain an AGN we adopted {\sevensize MAGAUTO} as the preferred magnitude measurement. \\

\noindent In Figure~\ref{fig:KS07} we show the significance as a function of the rest-frame UV magnitude (\filterr~band) for our sample. Only objects detected in all four epochs are considered. There are only $3$ objects with significance $S\ge3$, and hence displaying significant variability. This significance level is somewhat arbitrary since it is usually selected to keep the number of spurious detections low, as a trade off between purity and reliability. Since many of the objects of interest, i.e. confirmed AGN and \lyc~candidates, are below the significance limit, we also briefly investigate an alternative, more generous measure of variability in order to examine them more closely. We proceed as follows. A variability candidate is marked as ``possibly variable'' if its brightness deviates by $1\sigma\le\Delta<2\sigma$ from the $2008$ reference, where $\sigma=\sqrt{\sigma_X^2+\sigma_{2008}^2}$ is the combined photometric uncertainty, and $\Delta=(m_X-m_{2008})/\sigma_X$, for $X=2002$, $2003$, or $2007$. A candidate is marked as ``variable'' if its brightness deviates by $\Delta\ge2\sigma$ and the probability to randomly match the observed light curve is small. Sources with $\Delta<2\sigma$ are not considered to be variability candidates. \\

\noindent The variability profiles (light curves) for confirmed AGNs in our sample are shown in Figure~\ref{fig:varconfirmed}. We were unable to perform the variability test on the Xray AGN9 because it is below the detection limit in our \filterb~band data. To gain some idea of how significant the $\Delta$ variations are, we estimated how likely it is for a given deviation pattern to be randomly produced. The probability to obtain a random distribution of points that mimics the variability profiles we observe depends on the number of deviating points, the deviation sequence if deviating points are more than one,  the amplitude of the deviation, and the size of the individual uncertainty of each measurement.\\

\noindent The easiest way to account for all of these factors simultaneously is to perform a Monte Carlo simulation. For each source in Figure~\ref{fig:varconfirmed} we have created an array of $3$ random elements with the fourth element locked at $0$ (for the reference year $2008$), where each random data point was drawn from its individual Gaussian distribution with a standard deviation equal to the uncertainty of that data point. Thus created, if all of the random profile's data points or all of their errorbars enter the region defined by $\pm1\sigma$ of the original profile then this is considered a match to the original. For each object the total number of matches from $10000$ realizations is given as the percentage probability in Table~\ref{tab:var}. \noindent This random match probability is consistent with the significance parameter $S$ as shown in Figure~\ref{fig:pvss}.  \\

\noindent In summary, three out of the $14$ AGN (AGN5, AGN7, AGN8) have variability significance $S>3$ and random match probability $P\sim0.0\%$, all of them on the bright end with $M_R\leq-21.7$ AB. AGN11 has $S\sim2.5$, which is fairly high, but the random match probability is not negligible, $P\sim12\%$, and it is therefore not convincingly variable. The result of this test shows that we did not observe significant variability from low luminosity AGN ($M_R>-21.5$ AB). This implies that it is still possible for the \lyc~LAEs and LBGs from \papi~ to be hosting low luminosity AGNs below our variability detection limit. \\

\section{\lyc~and AGN properties}
\begin{figure}
\centering\tiny
\includegraphics[width=85mm]{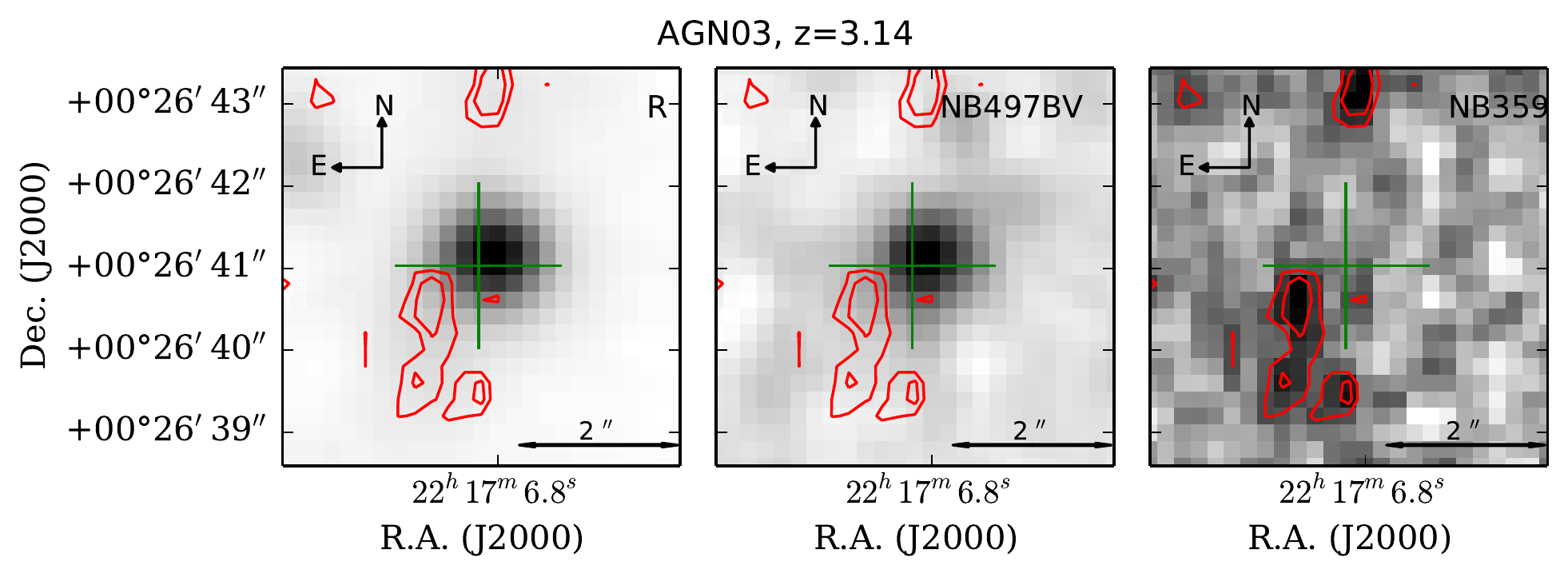}
\includegraphics[width=85mm]{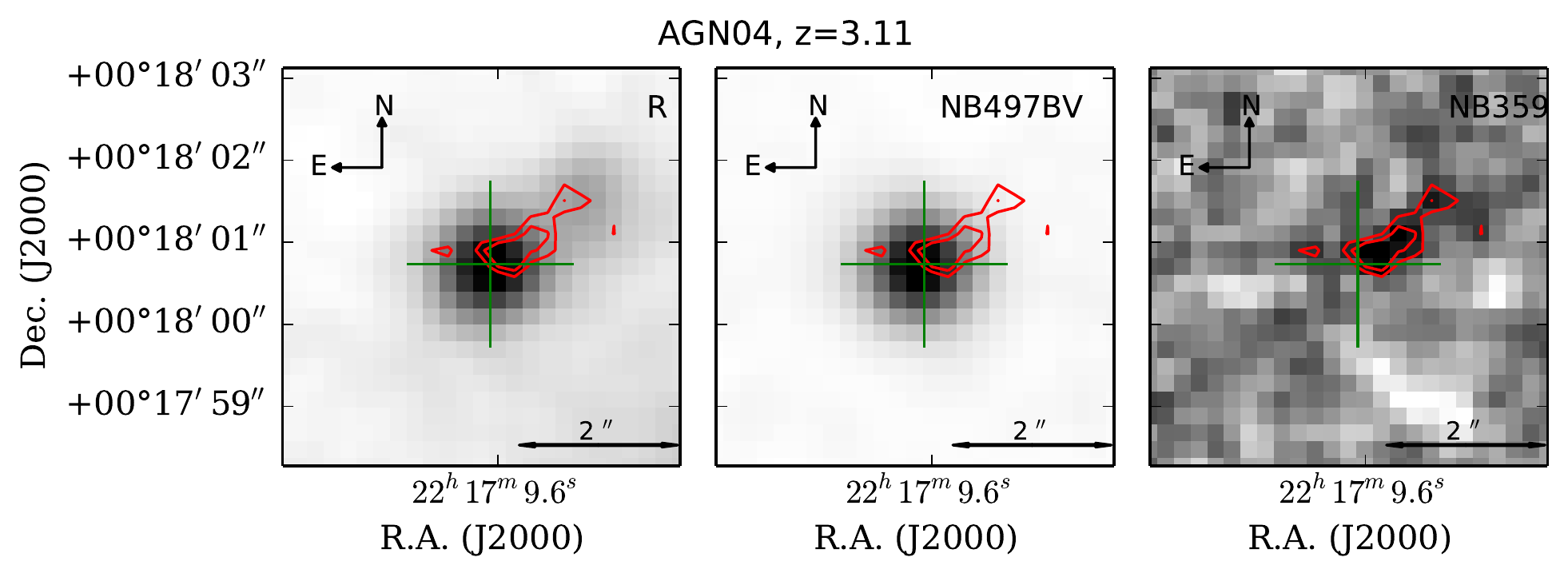}
\includegraphics[width=85mm]{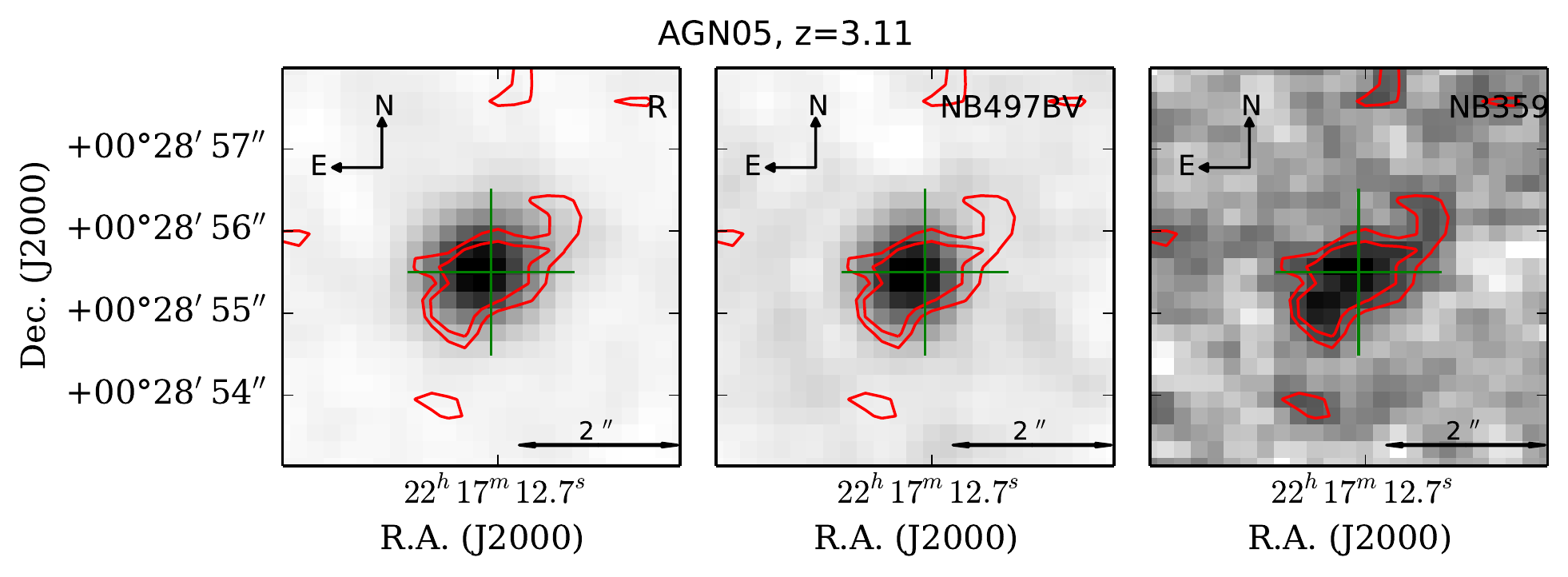}
\includegraphics[width=85mm]{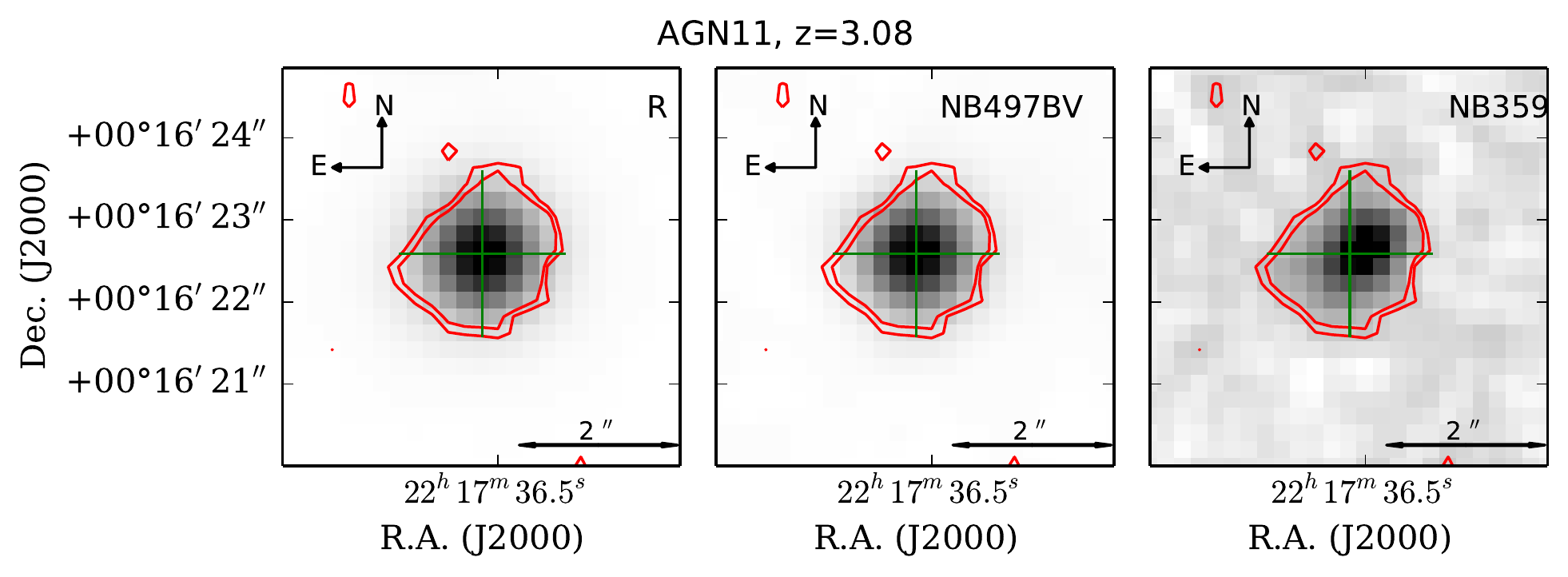}
\caption{AGN \lyc~candidates. The green cross marks the location of the \filterr~band detection. The filter order is \filterr, \filterlya$-$\filterbv~(continuum-subtracted~\lya), \filterlyc~(\lyc). $2$ and $3\sigma$ contours (red) trace the \lyc~detection in the \filterlyc~band, and are overplotted in all other bands. AGN3 is at a redshift which places the \lya~line at the edge of the filter, and so the \filterlya$-$\filterbv~image only marginally traces~\lya~emission. The contrast settings in each filter are global so that brightness comparison between the AGNs is possible. The scale bar is 2\arcsec~in all images. The color map is inverted. }\protect\label{fig:contours}
\end{figure}

\protect\label{sec:prop}
\noindent Figure~\ref{fig:contours} shows the four \lyc~AGNs (AGN3, AGN4, AGN5, AGN11) in our sample in the restframe UV continuum (\filterr~band, left panel), continuum-subtracted \lya~(\filterlya$-$\filterb\filterv, middle panel), and \lyc~emission (\filterlyc, right panel). We also present the complementary images of the AGNs with non-detections in Figure~\ref{fig:mosaic_non_detect} in the appendix. AGN3 and AGN4 are clearly morphologically different in \lyc~from AGN5 and AGN11, with the former two showing offsets between UV and \lyc, while UV and \lya~are well aligned. If the \filterlyc~detection is indeed \lyc~it can be coming from stars in offset starbursting regions. One could imagine a scenario in which non-stellar \lyc~could manifest an offset from the position of the AGN e.g. from dust scattering of the \lyc~from the accretion disk toward the direction of the observer, or from bound-free \lyc~emission from an ionized gas flow, though the latter would also scatter \lya~photons which we do not observe. The \lyc-emitting offset structure in AGN3 has $\diameter=1.2\arcsec$ fixed aperture magnitudes $\filterb=26.7$, $\filterv=25.9$, $\filterr=25.7$, $\filteri=26.0$, $\filterz=25.6$, $\filterlya=25.7$, and $\filterlyc=27.2$, with uncertainties $\le0.1$. Similarly, for AGN4 the magnitudes of the offset structure are $\filterb=26.6$, $\filterv=25.6$, $\filterr=25.4$, $\filteri=25.7$, $\filterz=25.6$, $\filterlya=24.2$, and $\filterlyc=27.1$. All of these are well above the $3\sigma$ limiting magnitude in the corresponding filter, except the \lyc~detections, which are close to the limiting magnitude of $\filterlyc (3\sigma)=27.4^{m}$.\\

\noindent The probability to have two or more foreground contaminants in our sample is low, $P(\ge2)=5.5\%$, however the probability for one of them to be a contaminant is not negligible, $P(=1)=26.7\%$. AGN3 is the most likely to be a contaminant because of its large spatial offset between \lyc~and UV, $\delta r=0.8$\arcsec. In Figure~\ref{fig:additional} in the appendix we show additional available data of the two AGNs manifesting offsets. These images reveal a complicated substructure for AGN3, spatially coincident with the \filterlyc~detection and visible across all of our multiwavelength data. If the substructure is at the same redshift as AGN3, it would be an indication of an ongoing merger. For AGN4 there is an additional distinctly separate object to the North-West (NW). The \lyc~detection seemingly originates from the center and stretches NW towards the second object, with no visible counterpart to the \lyc~bridge between the two objects in the available archival Hubble Space Telescope (HST) F814W image, although the data is not very deep. If these nearby objects truly associate with the AGN host galaxies their presence may be an indication of an ongoing merger or accretion of a smaller object which may have triggered the AGN in both AGN3 and AGN4. Hereafter we refer to AGNs $3$ and $4$ as stellar LyC AGNs for simplicity, although their \filterlyc~emission may come from contamination of foreground objects.\\

\begin{figure*}
\centering\tiny
\includegraphics[height=185mm]{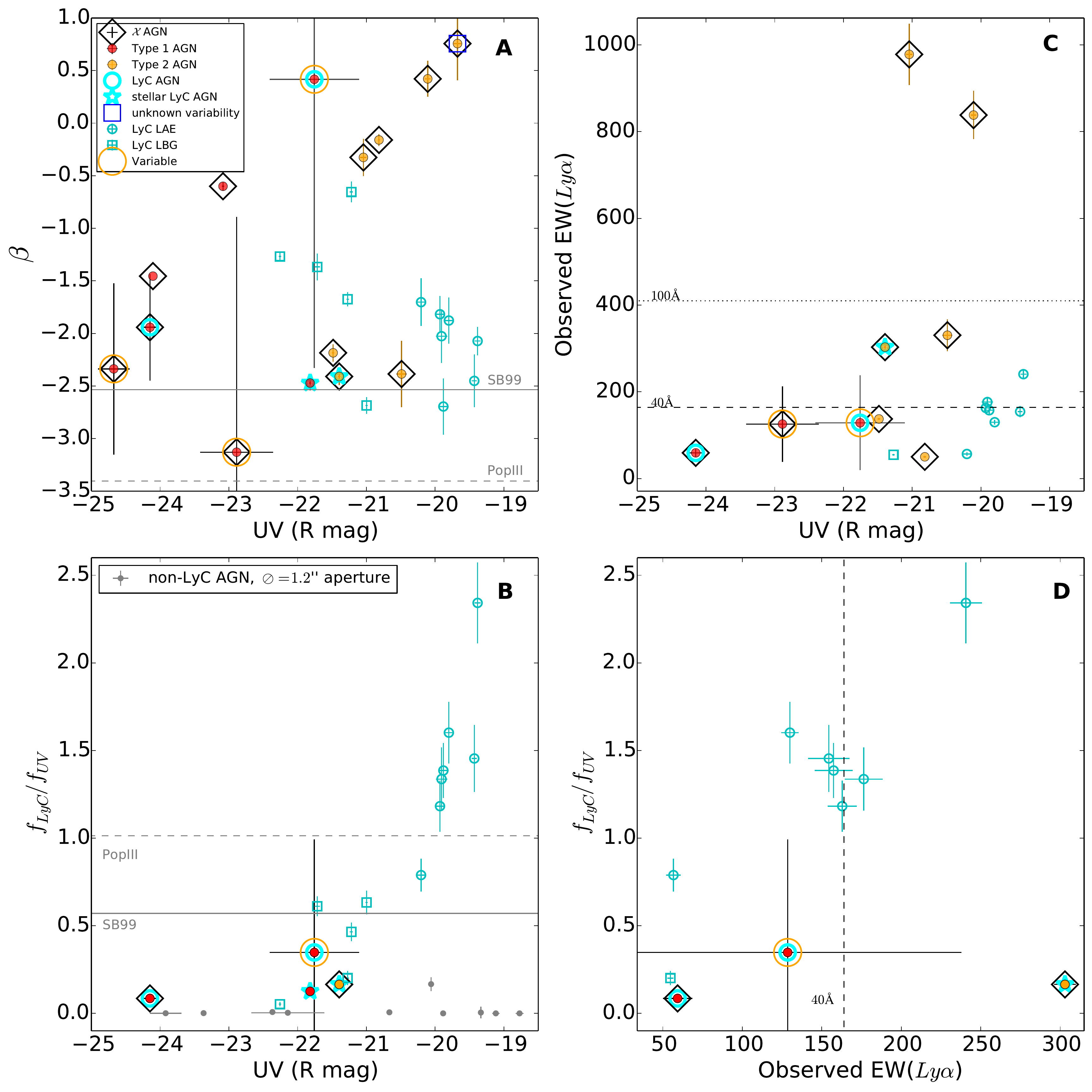}
\caption{Comparison between \lyc~AGN and the non-AGN \lyc~candidates from \papi. In panels A,B model predictions from {\sevensize STARBURST 99} (Salpeter IMF, slope $\alpha=2.35$, metallicity $Z=0.0004$, mass range $1$-$100$M${}_{\sun}$, age $0$ Myr), and {\sevensize PopIII} (\citet{2003A&A...397..527S}, same parameters as {\sevensize SB 99} except $Z=0.0$) are shown in solid, respectively dashed lines. Gray markers in panel B are from $1.2\arcsec$ aperture photometry of non-\lyc~AGN. Galactic extinction is applied to all measurements. \lyc~AGN3 is omitted from panels C,D because of its higher redshift of $z=3.14$. AGN4 (cyan star) is deviating from the rest in panel D because of the different nature of the emission (stellar in \filterlyc, coming from the AGN in all other filters). The rest-frame \ew$=40$\AA~and $100$\AA~are indicated for convenience.}\protect\label{fig:colormag}
\end{figure*}
\noindent If both or at least one of these AGNs is indeed showing stellar \lyc~escape the corresponding detection rates of $\sim0.14^{+0.33}_{-0.05}$ $(2/14)$ or $\sim0.07^{+0.24}_{-0.01}$ $(1/14)$ are comparable to \lyc~LAEs and LBGs at redshift $z\sim3.1$~\citep[e.g.][$6.6\%$-$13\%$]{2013ApJ...765...47N,2015arXiv150903996M}. The uncertainties on the detection rates are single-sided Poissonian upper and lower limits at a $68.4\%$ confidence level from~\citet{1986ApJ...303..336G}.\\


\begin{table*}
\begin{minipage}{186mm}
\caption{Photometry on the $14$ confirmed AGN at the position of the \filterr~band detections. Omitted uncertainties are $\leq0.1$. Uncertainties due to variability are not included.}
\protect\label{tab:phot}
\begin{tabular}{l c l l l l l l l l l r}
\hline
ID   &   FWHM  &  \filterr  &  \filterr($1.2\arcsec$) & \filterlyc & \filterlyc($1.2\arcsec$) & $(f_{1500}/f_{900})_{obs}$ & \filterb$-$\filterv& \filterlyc$-$\filterr & $\beta$ (UV slope) & $\delta r$\\
            AGN1 & $   0.8$ & $ 21.83$ & $ 22.57$ &                    &              &                      &  $  1.53        $&                & $-1.46 $        &\\
            AGN2 & $   0.8$ & $ 24.75$ & $ 25.52$ &                    &              &                      &  $  0.69        $&                & $-0.16 $        &\\
            AGN3 & $   0.9$ & $ 23.78$ & $ 24.58$ &  $ 26.04\pm 0.20$  &$ 28.11\pm 0.20$ &  $  7.97\pm 1.42$ &  $  1.18        $& $2.25\pm 0.20$ & $-2.47 $        & $0.8$\\
            AGN4 & $   0.9$ & $ 24.19$ & $ 25.00$ &  $ 26.14        $  &$ 27.00\pm 0.12$ &  $  6.06\pm 0.49$ &  $  1.01        $& $1.96        $ & $-2.41 $        & $0.5$\\
            AGN5 & $   0.9$ & $ 23.83$ & $ 24.62$ &  $ 24.98 $         &$ 26.29        $ &  $  2.88\pm 0.19$ &  $ -0.46        $& $1.15    $     & $ 0.42 $        & $0.2$\\
            AGN6 & $   0.9$ & $ 24.22$ & $ 25.03$ &                    &               &                     &  $  0.73        $&                & $-2.18 $        &\\
            AGN7 & $   0.8$ & $ 22.70$ & $ 23.44$ &                    &               &                     &  $ -0.18        $&                & $-3.13 $        & \\
            AGN8 & $   0.8$ & $ 21.04$ & $ 21.80$ &                    &               &                     &  $  1.62        $&                & $-2.34 $        & \\
            AGN9 & $   0.9$ & $ 25.91$ & $ 26.79$ &                    &               &                     &                  &                & $ 0.76\pm 0.35 $& \\
           AGN10 & $   0.9$ & $ 25.47$ & $ 26.27$ &                    &               &                     &  $  1.45        $&                & $ 0.42\pm 0.17$ & \\
           AGN11 & $   0.8$ & $ 21.42$ & $ 22.16$ &  $ 24.09$          & $ 24.96     $ &  $ 11.72\pm 0.27$   &  $  0.56        $& $2.67    $     & $-1.94 $        & $0.1$\\
           AGN12 & $   1.4$ & $ 25.09$ & $ 26.46$ &                    &               &                     &  $  1.12        $&                & $-2.39\pm 0.32 $&\\
           AGN13 & $   0.8$ & $ 22.69$ & $ 23.40$ &                    &               &                     &  $  1.02        $&                & $-0.60 $        &\\
           AGN14 & $   1.2$ & $ 24.53$ & $ 25.69$ &                    &               &                     &  $  0.87        $&                & $-0.33\pm 0.18 $&\\
\hline 
\end{tabular}
\end{minipage}
\medskip
Note: \filterlyc~samples the \lyc~regime at the relevant redshift of $z\ge3.06$. FWHM($\arcsec$) is measured from the reference \filterr~band. Galactic extinction by~\citet{2011ApJ...737..103S} has been applied to all values. The $(f_{1500}/f_{900})$ column is the observed total flux density ratio of non-ionizing to ionizing radiation. $\beta$ is the UV slope estimated from \filterv$-$\filteri~for all AGN, except for AGN1, AGN8, AGN13 for which \filterr$-$\filterz~was used. $\delta r$ ($\arcsec$) is the \filterr$\leftrightarrow$\filterlyc~offset in centroid position, with an astrometric uncertainty of $\sim0.2$\arcsec.
\end{table*}
\noindent The \lyc~detection rate for the Type I AGNs in our sample is $0.29^{+0.66}_{-0.10}$ $(2/7)$, where we exclude stellar \lyc~AGN3. Compared to \lyc~LAEs and LBGs at the same redshift this is higher by a factor of $2.2^{+7.3}_{-1.5}$. Due to the small number statistics we are unable to claim this difference is significant. Similar to~\citet{2009ApJ...692.1476C} the detection rate among Type II AGN is $0\%\,(0/7)$, where we exclude stellar \lyc~AGN4. Among Xray AGN the detection rate is $0.09^{+0.30}_{-0.02}$ $(1/11)$, comparable to \lyc~LAEs and LBGs at the same redshift. Among protocluster members the detection rate is $0.20^{+0.46}_{-0.07}$ $(2/10)$. We note here that regardless of which group of AGN we consider, the rate of detection is a far cry from $100\%$.\\ 

\noindent In the four panels of Figure~\ref{fig:colormag} we examine any connection between variability, presence of Xrays, AGN type, and the properties of the \lyc~AGN compared to the viable \lyc~LAEs and LBGs candidates from \papi~(none of which are AGN), using photometry from Table~\ref{tab:phot}. Confirmed and possible contaminants among \lyc~LAEs and LBGs are excluded from this comparison. We estimate the UV slope $\beta$ from \filterv$-$\filteri~for all AGN, except for the higher redshift AGN1, AGN8, and AGN13, for which we use \filterr$-$\filterz to avoid having emission lines in the filters. \\

\noindent Since our multiband data were not taken at the same epoch, with e.g. \filterr~in 2001 and \filterlyc~in 2008, for AGN5, AGN7, AGN8, and AGN11 we add the uncertainty due to variability calculated as the dispersion of the light curves in Figure~\ref{fig:varconfirmed}. This dispersion amounts to $0.65^{m}$, $0.53^{m}$, $0.23^{m}$, and $0.12^{m}$ respectively. First we note that all Type II AGN are on the faint end ($\gtrsim-21.5^{m}$), while the majority of the Type I AGN have quasar-like luminosities ($\lesssim-23^{m}$). One out of two \lyc~AGN is variable, and both stellar \lyc~AGN are non-variable, suggesting that significant variability has no detectable influence on \lyc~leakage. There also seems to be no connection between the presence of Xrays and the \lyc~leakage in AGN in terms of either rest-frame UV slope or ionizing to non-ionizing flux density ratio (panels~\ref{fig:colormag}A and~\ref{fig:colormag}B). \lyc~LAEs and \lyc~LBGs have rest-frame UV slopes similar to the \lyc~AGN, and \lya~equivalent widths that are comparably large (panels~\ref{fig:colormag}A and~\ref{fig:colormag}C). In terms of the flux density ratio $f_{LyC}/f_{UV}$ the \lyc~AGN are comparable to \lyc~LBGs, and much lower than \lyc~LAEs (panel \ref{fig:colormag}B). The ratio of the latter is notably $>1$, which is not consistent with purely stellar \lyc~emission and would require nebular \lyc~contribution (see \papi~for more details on the \lyc~LAEs). The \lyc~flux density ratio versus the \ew~in panel D shows that \lyc~AGN are consistent with the tentatively suggested correlation between \lya~and \lyc~in \papi. Stellar \lyc~AGN4 is deviating in this figure due to the nature of the radiation being stellar in \filterlyc~but coming from the AGN in all other filters. To obtain a correct (stellar) flux density ratio for this object one should subtract the AGN in the center and remeasure the photometry of the residual. We attempted to do this with GALFIT however our resolution is $FWHM\sim1\arcsec$ which corresponds to the size of the object and the residual we obtain shows no detection above $3\sigma$.\\

\noindent In panel C two AGNs show very high observed equivalent widths, AGN10 with \ew$\,\sim840$\AA~and AGN14 with \ew$\,\sim980$\AA. These two objects were classified as AGNs in~\citet{2015MNRAS.450.2615S} due to superimposed Xray emission. We examined the SEDs of these two objects (Figure~\ref{fig:hostdominated} in the appendix) and the imaging data in all filters. The SEDs reveal spectra redder than the typical quasar spectrum, which may indicate that they are obscured or even host-dominated. The continuum-subtracted \lya~images (Figure~\ref{fig:mosaic_non_detect}) show extended emission, which is expected for AGN10, classified as a Lyman $\alpha$ Blob (LAB) by~\citet{2005MNRAS.363.1398G}. AGN14 has not previously been identified as a LAB, however Figure~\ref{fig:mosaic_non_detect} shows it to be even more extended than AGN10, with a clealy visible large \lya~halo.\\

\begin{figure*}
\centering\tiny
\includegraphics[height=53mm]{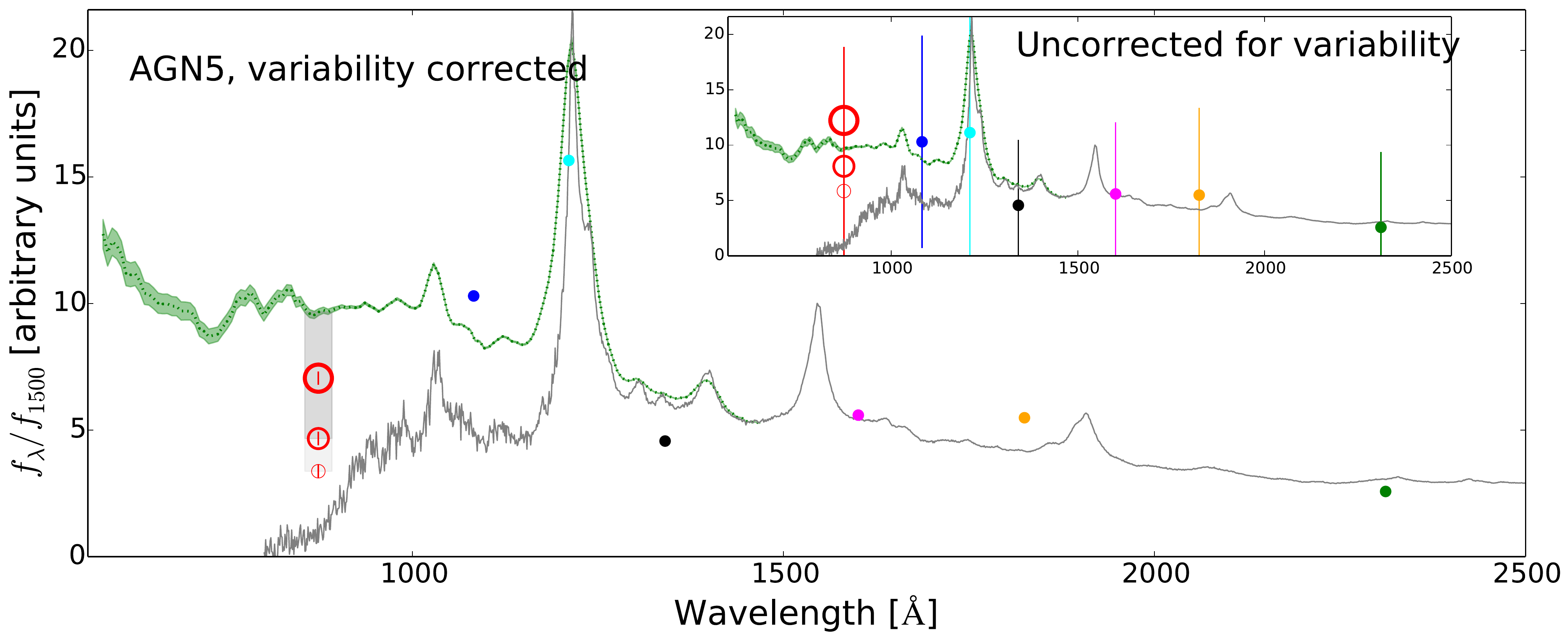}
\includegraphics[height=53mm]{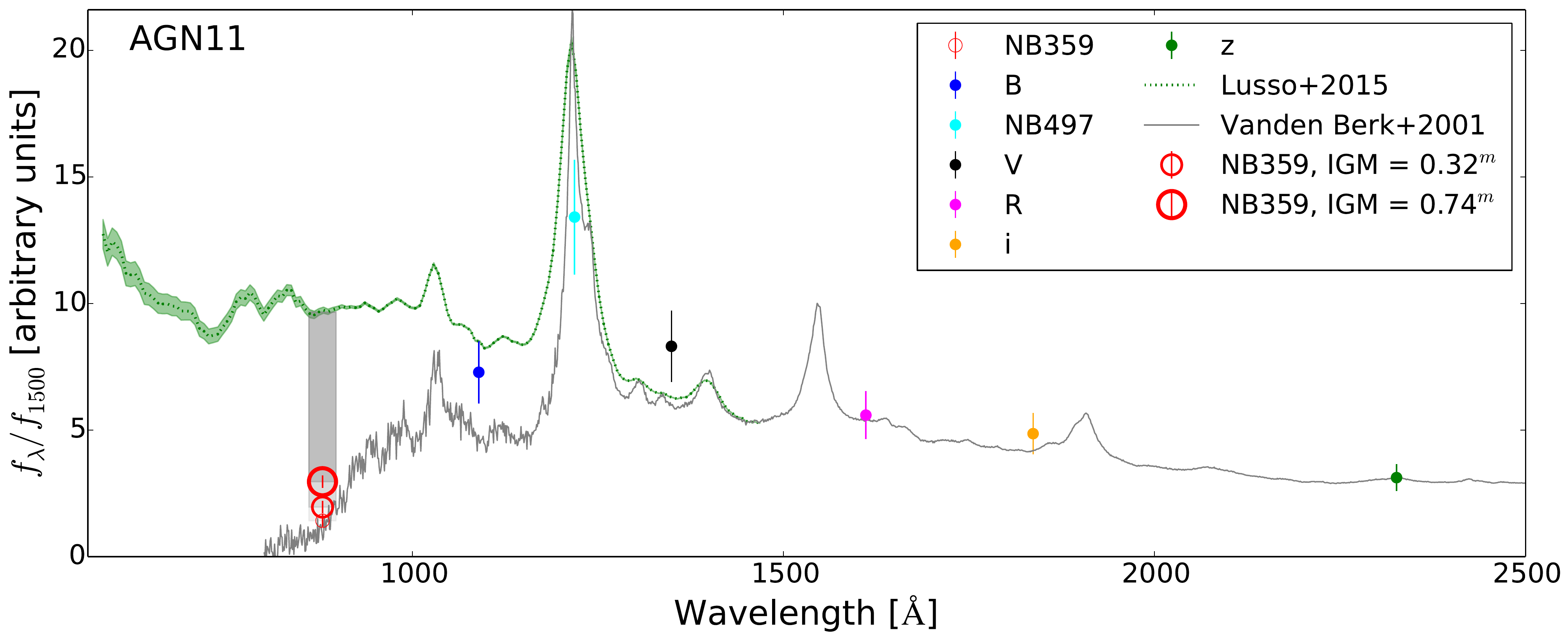}
\caption{SEDs of the \lyc-leaking AGN scaled to match the standard~\citet{2001AJ....122..549V} spectrum in the \filterr~band. We assume the IGM corrected stacked spectrum by~\citet{2015MNRAS.449.4204L} gives a good representation of $f^{\lyc}_{intr}$ and use it to estimate the escape fraction at the wavelength of the \filterlyc~band. Gray rectangles indicate the rest-frame width of the \filterlyc~band (horizontal) and the extent of $f^{\lyc}_{obs}$ to $f^{\lyc}_{intr}$ (vertical) used to obtain $f_{esc}$ in Table~\ref{tab:fesc}. The inset figure for AGN5 is the uncorrected SED, with large uncertainties due to significant variability (see text for details).}\protect\label{fig:seds}
\end{figure*}

\noindent Figure~\ref{fig:seds} shows the spectral energy distribution (SED) of the two \lyc~AGN, which are fairly consistent with the~\citet{2001AJ....122..549V} stacked spectrum at $\lambda>1000$\AA. \lyc~AGN5 is strongly variable, with the multiband observations taken at different epochs. In the inset figure we add in quadrature to the formal photometry error of each band the uncertainty due to variability, estimated from the dispersion of the light curve in Figure~\ref{fig:varconfirmed} and amounting to $0.65^{m}$. AGN11 is marginally variable, but we add the dispersion of its light curve, which is $0.12^m$. It is difficult to interpret a flux density ratio of a strongly variable object when the data are not taken in the same epoch. Without correcting the SED for variability applying a median IGM attenuation ($\tau=0.74^{m}$, \citet{2014MNRAS.442.1805I}) to the \filterlyc~data point overshoots the IGM corrected spectrum, which we assume to be a good representation of the intrinsic spectrum. We therefore apply a variability correction to AGN5 by assuming the \filterb~band light curve in Figure~\ref{fig:varconfirmed} is representative of the light curves in all other filters. The correction brings all filter measurements to the level of the reference year $2008$. The \filterlyc~image has no correction since it was taken in the year $2008$. The \filterr,~\filteri, and \filterz~bands were taken in the years $2000$ and $2001$, which are not covered by the \filterb~band light curve. The precise correction would be the amplitude between the magnitude in $2008$ and the year of observation, however since we do not have \filterb~band data in $2000$, $2001$ we assign the average of the light curve $\overline{m}$ to these years and calculate their correction amplitude as $A=m(2008)-\overline{m}$. The correction resulted in brightening the \filterr,~\filteri, and \filterz~fluxes by $0.60^{m}$. The \filterb~and \filterv~bands are average stacks over the range of the light curve, so they were corrected by the same amount. The \filterlya~image was taken in $2002$ so it was corrected by the full amplitude $A=m(2008)-m(2002) = -0.97^{m}$.\\

\noindent Even with the variability correction the SED shows an excess in the \filterb~band (blue circle) compared to the~\citet{2001AJ....122..549V} spectrum and is instead more consistent with the intrinsic~\citet{2015MNRAS.449.4204L} spectrum. The broad {\sevensize OVI} emission line falls well inside the \filterb~band filter (Figure~\ref{fig:spectra}), however from a Gaussian fit to the line we estimate that it contributes $\sim20\%$ to the \filterb~band flux. Although the B band suffers from IGM attenuation much less than the \filterlyc~(\lyc) band, this may suggest an almost transparent line of sight to AGN5.

\begin{table*}
\begin{minipage}{184mm}
\centering
\caption{Average flux density ratio and ionizing emissivity $\epsilon_{LyC}$ [$10^{24}$ erg s${}^{-1}$Hz${}^{-1}$Mpc${}^{-3}$] at $z\sim3.1$ from direct measurement for two magnitude ranges $(-18.0,-28.0)$ and $(-20.0,-28.0)$. The \filterz~band samples the UV continuum. All values have been statistically corrected for foreground contamination. Non-detections are given as $3\sigma$ upper limits. The uncertainties in the $\left<\textit{z}'\right>$ and $\left<\textit{V}-\textit{i}'\right>$ columns are $<0.03$, $<0.05$ respectively. The errors on $\epsilon_{LyC}$ include luminosity function uncertainties. $\mathbb{N}$ is the number of stacked objects.}\protect\label{tab:stackave}
\begin{tabular}{ l c c c c c c }
\hline
\multicolumn{6}{c}{CASE A}\\
\multicolumn{6}{c}{Full sample, no variability correction}\\
\multicolumn{6}{c}{max sample size = 13}\\
Stack             & $\left< f_{LyC} / f_{UV} \right>$  & $\left<\textit{z}'\right>$ & $\left<\textit{V}-\textit{i}'\right>$ & $\epsilon_{LyC}(-18.0,-28.0)$& $\epsilon_{LyC}(-20.0,-28.0)$&$\mathbb{N}$ \\
\hline
               all  & $<0.045       $  & $23.50$  &  $0.05$  & $<0.481       $ & $<0.423       $  &$=13$     \\ 
              \lyc  & $0.109\pm0.012$  & $23.01$  &  $0.10$  & $1.164\pm0.132$ & $1.025\pm0.117$  &$\leq3$  \\ 
             TypeI  & $0.051\pm0.006$  & $22.75$  &  $0.04$  & $0.546\pm0.068$ & $0.481\pm0.060$  &$\leq7$  \\ 
   TypeI$\cap$Xray  & $<0.033       $  & $22.61$  &  $0.01$  & $<0.352       $ & $<0.310       $  &$\leq5$  \\ 
              Xray  & $<0.057       $  & $23.45$  &  $0.04$  & $<0.606       $ & $<0.533       $  &$\leq11$ \\ 
  noLyC$\cap$TypeI  & $<0.009       $  & $22.59$  &  $0.01$  & $<0.099       $ & $<0.087       $  &$=3$      \\ 
   noLyC$\cap$Xray  & $<0.059       $  & $23.62$  &  $0.05$  & $<0.634       $ & $<0.559       $  &$=9$      \\ 
\hline
\multicolumn{6}{c}{CASE B}\\
\multicolumn{6}{c}{stellar LyC excluded, no variability correction}\\
\multicolumn{6}{c}{max sample size = 11}\\
Stack             & $\left< f_{LyC} / f_{UV} \right>$  & $\left<\textit{z}'\right>$ & $\left<\textit{V}-\textit{i}'\right>$ & $\epsilon_{LyC}(-18.0,-28.0)$& $\epsilon_{LyC}(-20.0,-28.0)$\\
\hline
               all  & $<0.055       $  & $23.37$  &  $0.07$  & $<0.586       $ & $<0.516       $  &$=11$      \\ 
              \lyc  & $0.139\pm0.015$  & $22.84$  &  $0.20$  & $1.489\pm0.160$ & $1.311\pm0.141$  &$\leq2$   \\
             TypeI  & $0.049\pm0.006$  & $22.61$  &  $0.04$  & $0.527\pm0.069$ & $0.464\pm0.061$  &$\leq5$   \\
   TypeI$\cap$Xray  & $0.016\pm0.003$  & $22.43$  &  $0.02$  & $0.169\pm0.033$ & $0.149\pm0.029$  &$\leq4$   \\
              Xray  & $<0.058       $  & $23.36$  &  $0.04$  & $<0.618       $ & $<0.545       $  &$\leq10$  \\ 
  noLyC$\cap$TypeI  & $<0.013       $  & $22.59$  &  $0.01$  & $<0.138       $ & $<0.121       $  &$=3$       \\ 
   noLyC$\cap$Xray  & $<0.069       $  & $23.62$  &  $0.05$  & $<0.740       $ & $<0.652       $  &$=9$       \\ 
\hline
\multicolumn{6}{c}{CASE C}\\
\multicolumn{6}{c}{stellar LyC excluded, variability corrected}\\
\multicolumn{6}{c}{max sample size = 11}\\
Stack             & $\left< f_{LyC} / f_{UV} \right>$  & $\left<\textit{z}'\right>$ & $\left<\textit{V}-\textit{i}'\right>$ & $\epsilon_{LyC}(-18.0,-28.0)$& $\epsilon_{LyC}(-20.0,-28.0)$\\
\hline
               all  & $<0.053       $  & $23.62$  &  $0.05$  & $<0.570       $ & $<0.502       $  &$=11$      \\ 
              \lyc  & $0.092\pm0.009$  & $22.60$  &  $0.18$  & $0.981\pm0.094$ & $0.864\pm0.083$  &$\leq2$   \\
             TypeI  & $0.035\pm0.004$  & $22.57$  &  $0.03$  & $0.372\pm0.042$ & $0.327\pm0.037$  &$\leq5$   \\
   TypeI$\cap$Xray  & $0.015\pm0.002$  & $22.45$  &  $-0.01$ & $0.166\pm0.026$ & $0.146\pm0.023$  &$\leq4$   \\
              Xray  & $<0.060       $  & $23.39$  &  $0.01$  & $<0.641       $ & $<0.564       $  &$\leq10$  \\ 
  noLyC$\cap$TypeI  & $<0.009       $  & $23.39$  &  $-0.04$ & $<0.099       $ & $<0.087       $  &$=3$       \\ 
   noLyC$\cap$Xray  & $<0.065       $  & $23.39$  &  $0.00$  & $<0.698       $ & $<0.615       $  &$=9$       \\ 
\hline
\multicolumn{6}{c}{CASE D}\\
\multicolumn{6}{c}{stellar LyC and variable AGN excluded}\\
\multicolumn{6}{c}{max sample size = 8}\\
Stack             & $\left< f_{LyC} / f_{UV} \right>$  & $\left<\textit{z}'\right>$ & $\left<\textit{V}-\textit{i}'\right>$ & $\epsilon_{LyC}(-18.0,-28.0)$& $\epsilon_{LyC}(-20.0,-28.0)$\\
\hline
               all  & $<0.077       $  & $23.66$  &  $0.18$  & $<0.827       $ & $<0.728       $  & $=8$      \\ 
      \lyc (AGN11)  & $0.062\pm0.003$  & $21.94$  &  $0.03$  & $0.661\pm0.029$ & $0.582\pm0.026$  & $\leq1$  \\
             TypeI  & $0.031\pm0.003$  & $22.41$  &  $0.21$  & $0.334\pm0.033$ & $0.294\pm0.029$  & $\leq2$  \\
   TypeI$\cap$Xray  & $0.031\pm0.004$  & $22.42$  &  $0.22$  & $0.327\pm0.039$ & $0.288\pm0.034$  & $\leq2$  \\
              Xray  & $<0.078       $  & $23.80$  &  $0.24$  & $<0.833       $ & $<0.734       $  & $\leq8$  \\ 
  noLyC$\cap$TypeI  & $<0.019       $  & $22.88$  &  $0.58$  & $<0.205       $ & $<0.181       $  & $=1$      \\ 
   noLyC$\cap$Xray  & $<0.084       $  & $24.46$  &  $0.50$  & $<0.894       $ & $<0.787       $  & $=7$      \\ 
\hline
\multicolumn{6}{c}{CASE E}\\
\multicolumn{6}{c}{Type I AGN at $z\sim3.1$}\\
\multicolumn{6}{c}{max sample size = 4}\\
Stack             & $\left< f_{LyC} / f_{UV} \right>$  & $\left<\textit{z}'\right>$ & $\left<\textit{V}-\textit{i}'\right>$ & $\epsilon_{LyC}(-18.0,-28.0)$& $\epsilon_{LyC}(-20.0,-28.0)$\\
\hline
             TypeI  & $0.060\pm0.009$  & $22.96$  &  $-0.04$  & $0.646\pm0.099$ & $0.569\pm0.087$ & $\leq4$ \\
\hline 
\end{tabular}
\end{minipage}
\end{table*}

\section{Average flux density ratios}\protect\label{sec:averatio}
\noindent The observed average flux density ratio from the AGNs in our sample is obtained through stacking and shown in Table~\ref{tab:stackave}. Since the sample contains Type I AGNs, which can have a broad {\sevensize CIV} emission line falling inside of the \filterr~band filter, here we use the \filterz~band data to measure the UV continuum. All stacks were statistically corrected for foreground contaminantion in a Monte Carlo (MC) simulation of 100 realizations, using the PMF of expected foreground contaminants in the same manner as in \papi. Briefly, we make $10\arcsec\times10\arcsec$ cutouts of \filterlyc~for each object and normalize them by the $\diameter=1.2\arcsec$ aperture flux in the \filterz~band. In each MC run we draw a random number of contaminants $n_{cont}$ based on the PMF of expected number of foreground contaminants, with a separate PMF obtained for each sample size. If the total number of candidates is $n$ and the number of non-detections in \filterlyc~$N$, we make a stack with $N+n-n_{cont}$ and measure the flux density ratio inside a $\diameter=1.2\arcsec$ aperture. If $n_{cont}\ge n$ then only $N$ number of images are stacked. If the stack consists only of \lyc~candidates the flux density ratio is set to zero for that MC run. The result is the average value of $100$ such realizations. We estimate the uncertainty of the flux density ratio in the following way. For the \textit{i}th object in the \textit{j}th realization we sample $100$ random sky positions in the \filterlyc~band, normalize each by object \textit{i}'s \filterz~band flux, and measure the aperture sums. The uncertainty per object, $\sigma_{i,j}$, is the standard deviation of these $100$ aperture sums. The uncertainty per average stack in MC run \textit{j} is then $\sigma_{stack,j}=\sqrt{\sum{\sigma_{i,j}^{2}}}/\mathbb{N}$, $i=1\ldots \mathbb{N}$ where $\mathbb{N}=N+n-n_{cont}$ is the total number of objects stacked. The final uncertainty of the flux density ratio we present in Table~\ref{tab:stackave} is the average of $100$ MC runs.\\

\noindent We considered five cases, shown in Table~\ref{tab:stackave}. In all cases we exclude AGN1 because due to its higher redshift ($z=3.8$) the \filterz~band covers a different rest-frame wavelength range than the rest of the sample. Case A takes the sample of $13$ AGNs, including the two stellar \lyc~AGNs. Case B excludes the two stellar \lyc~AGNs and AGN1. Case C is like B but applies a variability correction to the photometry of the three variable AGNs. Case D is like B but considers only non-variable AGNs. Case E considers only a stack of the four Type I AGN at redshifts $3.084\leq z\leq3.140$. Cases have subgroups consisting of all objects, only \lyc~candidates, all Type I AGNs, all Xray AGNs, all Type I AGNs with Xray, all Type I AGNs with non-detections in \filterlyc, and all Xray AGNs with non-detections in \filterlyc. For case D the subgroup of only \lyc~candidates contains just one object, AGN11. Table~\ref{tab:stackave} also shows $\left<\textit{z}'\right>$ and $\left<\textit{V}-\textit{i}'\right>$ which sample the UV continuum and slope.\\

\noindent The variability corrections for Case C were obtained for all three variables as described in Section~\ref{sec:prop}, namely we assign an average magnitude $\overline{m}_B$ to the missing $2001$ data point from the \filterb~band light curve, measure the amplitude $A=m_B(2008)-\overline{m}_B$ and use it to correct the \filterv\filteri\filterz~fluxes. The corrections brighten these fluxes by $0.60^{m}$ and $0.62^{m}$ for AGN5 and AGN7, respectively, and dim the fluxes by $0.21^{m}$ for AGN8.

\section{Ionizing Emissivity}\protect\label{sec:emissivity}
\noindent The literature often assumes an average escape fraction of unity and a double power law spectrum to estimate the ionizing emissivity of high-z AGNs. Since we have direct measurements of the \lyc-to-UV flux density ratios from two AGN in our sample, we could instead compute the observed ionizing emissivity, and compare it to results which assume a unity escape fraction. We will refer to the emissivity obtained from direct \lyc~measurement as $\epsilon_{\lyc}$, the emissivity from assuming a double power law spectrum with $f_{esc}=1$ as $\epsilon_{912}$, and the emissivity inferred from Ly$\alpha$ forest observations as $\epsilon_{912}^{L\alpha F}$. Throughout this section the emissivity is in comoving units of $10^{24}$ erg s${}^{-1}$ Hz${}^{-1}$ Mpc${}^{-3}$. To calculate $\epsilon_{\lyc}$ we use the quasar luminosity function from~\citet{2012ApJ...755..169M} at redshift $z=3.2$. This is described by a double power-law function with parameters $\phi_{\star}=2.65(\pm2.22)\cdot10^{-7}$ Mpc${}^{-3}$ mag${}^{-1}$, $\alpha=-2.98\pm0.21$, $\beta=-1.73\pm0.11$, and $M_{\star}=-25.54\pm0.68$ at $\lambda=1450$\AA. Since we are sampling the UV continuum at the position of the \filterz~band ($\lambda_{\textrm{eff}}=2317$\AA), we scale $M_{\star}^{1450}$ to $M_{\star}^{2317}$ using the~\citet{2015MNRAS.449.4204L} spectrum. The non-ionizing emissivity is then given by
\begin{equation}
\begin{split}
\epsilon_{UV}=\int{\phi(L_{2317,z})L_{2317}\mathrm{d}L_{2317}}
\end{split}
\end{equation}
\noindent where $\phi$ is the luminosity function, $L_{2317}$ is the non-ionizing UV continuum luminosity. For the magnitude range of the integration $(-18,-28)$ and $(-20,-28)$, the resulting non-ionizing emissivity is $\epsilon_{UV,24}=5.111$, respectively $4.501$, where the subscript $24$ means $10^{24}$ erg s${}^{-1}$ Hz${}^{-1}$ Mpc${}^{-3}$. The ionizing emissivity can be calculated as $\epsilon_{\lyc}=\mathcal{R}\epsilon_{UV}$, where $\mathcal{R}$ is the flux density ratio $f_{LyC}/f_{UV}e^{\tau_{IGM}}$ and may be luminosity dependent as seen in Figure~\ref{fig:colormag}B. For the observed average flux density ratio we use the direct detections or the $3\sigma$ upper limits in Table~\ref{tab:stackave} for the four cases we considered. The IGM attenuation is assumed to be $\tau_{IGM}=0.74^{m}$, which is the median of the MC simulation in the new IGM transmission model in~\citet{2008MNRAS.387.1681I} but with an updated IGM absorbers’ statistics described in~\citet{2014MNRAS.442.1805I}. Table~\ref{tab:stackave} shows the resulting ionizing emissivities for all cases and subgroups. The error budget in $\epsilon_{\lyc}$ includes the flux density ratio and luminosity function uncertanties, added in quadrature.\\

\noindent Many $\epsilon_{\lyc}$ reported in the literature assume a double power law and $f_{esc}=1$. It is immediately obvious from our data that the average escape fraction is most likely not unity. To further illustrate this we calculate the ionizing emissivity assuming a double power law spectrum following~\citet{2015A&A...578A..83G}, with $f_{\lambda}\propto\lambda^{\alpha}$ with $\alpha=1.57$ for $\lambda\leq1200$\AA, and $f_{\lambda}\propto\lambda^{\beta}$ with $\beta=0.44$ for $\lambda>1200$\AA. It is equally interesting to compare our $\epsilon_{\lyc}$ from direct measurements to the value obtained using the~\citet{2015MNRAS.449.4204L} IGM corrected spectrum, which has steeper values of $\alpha=1.70$ for $\lambda\leq912$\AA~and $\beta=0.61$ for $\lambda>912$\AA. These two equations are

\begin{equation}\protect\label{eq:telfer}
\epsilon_{912}=\int{\phi(L_{2317,z})L_{2317}\left(\frac{912}{1200}\right)^{1.57}\left(\frac{1200}{2317}\right)^{0.44}\mathrm{d}L_{2317}}
\end{equation}
\begin{equation}\protect\label{eq:lusso}
\epsilon_{912}=\int{\phi(L_{2317,z})L_{2317}\left(\frac{912}{2317}\right)^{0.61}\mathrm{d}L_{2317}}
\end{equation}

\noindent With Eq.~\ref{eq:telfer} (Giallongo) the ionizing emissivity becomes $\epsilon_{912,24}=2.487$ and $2.190$, respectively for the magnitude ranges $(-18,-28)$ and $(-20,-28)$. With Eq.~\ref{eq:lusso} (Lusso) the values are $\epsilon_{912,24}=2.894$ and $2.549$. The largest value for $\epsilon_{\lyc,24}$ in Table~\ref{tab:stackave} is obtained for the subgroup of only \lyc~AGN candidates (Case C, variability corrected), $\epsilon_{\lyc,24}=0.981\pm0.094$ for $(-18,-28)$. This group is not representative of all AGN and may be biased towards high escape fractions. Even so, the ionizing emissivity from direct measurement $\epsilon_{LyC,24}$ is lower than the $f_{esc}=1$ case $\epsilon_{912,24}$ values by a factor of $\sim2.5$. For the more representative Type I subgroup it is a factor of $\sim6.7$ lower. The discrepancy increases for Case D (non-variable AGN). Even if we only look at Type I AGNs strictly at $z\sim3.1$ (Case E), $\epsilon_{LyC,24}$ is a factor of $\sim3.8$ lower than $\epsilon_{912,24}$. We note that $\epsilon_{LyC,24}$ for Case B is larger than for all other cases, being only a factor of $\sim1.7$ lower than $\epsilon_{912,24}$. However, we remind the reader that this case contains no variability correction and the \lyc~and UV continua are sampled at different time epochs, which makes their ratio hard to interpret.\\

\begin{figure}
\centering\tiny
\includegraphics[width=85mm]{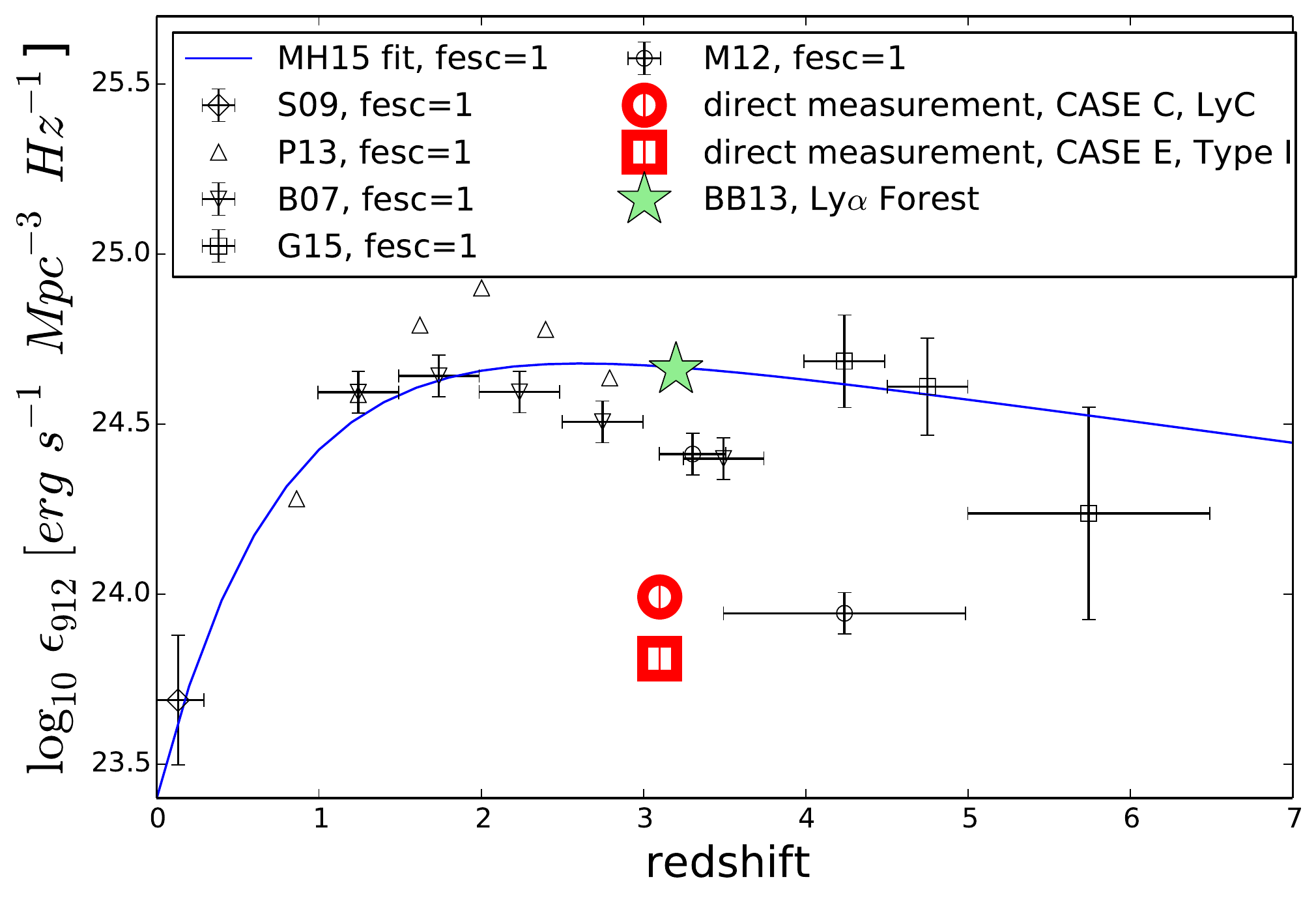}
\caption{Ionizing emissivity vs. redshift, reproduced from~\citet[][MH15]{2015ApJ...813L...8M}. The~\citet[][BB13]{2013MNRAS.436.1023B} emissivity at $z=3.2$ inferred from Ly$\alpha$ forest observations in quasar spectra is shown with a green star symbol. Our direct measurement of $\epsilon_{LyC}$ (thick red markers) is given for the largest but biased value (Case B, \lyc~subgroup), and for Type I AGN between $3.084\leq z\leq3.140$ (Case E). Literature results (open black markers) compiled by MH15, all inferred from quasar luminosity functions assuming $f_{esc}=1$, are shown from~\citet[][S09]{2009A&A...507..781S},~\citet[][P13]{2013A&A...551A..29P},~\citet[][B07]{2007A&A...472..443B},~\citet[][G15]{2015A&A...578A..83G},~\citet[][M12]{2012ApJ...755..169M}. The line is the MH15 fit through the $f_{esc}=1$ literature data. All literature data has been re-scaled to $M_{\star}^{2317}$.}\protect\label{fig:emissivity}
\end{figure}

\noindent Our direct measurement $\epsilon_{\lyc,24}$ is sampled at $\lambda\sim876$\AA~instead of at the Lyman edge $912$\AA, but the ratios $\epsilon_{912}/\epsilon_{876}$ for both double power laws are very close to unity, namely $\epsilon_{912}/\epsilon_{876}=1.08$ for Eq.~\ref{eq:telfer} and $\epsilon_{912}/\epsilon_{876}=1.04$ for Eq.~\ref{eq:lusso}. The observed difference between direct measurement of \lyc~and assuming a power law can therefore not be explained by the wavelength region at which \lyc~is being sampled. \\

\noindent In Figure~\ref{fig:emissivity} we reproduce the emissivity vs. redshift plot of~\citet{2015ApJ...813L...8M}, who compile quasar luminosity function data from the literature, convert the integrated optical emissivity using the new~\citet{2015MNRAS.449.4204L} stacked spectrum, assume average $f_{esc}=1$, and infer the ionizing emissivity for different redshifts. We overplot our emissivity $\epsilon_{\lyc}$ from direct detections for the largest value in Table~\ref{tab:stackave} (Case C, \lyc~subgroup) and for the Type I AGNs at redshift $z\sim3.1$ (Case E). Both fall clearly far beneath the $f_{esc}=1$ emissivities.~\citet{2013MNRAS.436.1023B} infer the ionizing emissivity at $z=3.2$ from observations of the Ly$\alpha$ forest in composite quasar spectra. They obtain $\epsilon_{912,24}^{L\alpha F}=8.147$. Our $\epsilon_{\lyc}$ from direct measurement fall obviously short of sustaining such levels and further appear to not dominate the ionization budget at this redshift since they only make up at most $12\%$ (Case C, \lyc~subgroup) and as little as $\sim5\%$ (Case E) of $\epsilon_{912,24}^{L\alpha F}$. This may imply that faint AGN, even if significantly more numerous at high redshifts than previously thought~\citep{2015A&A...578A..83G}, may not contribute enough to $f_{esc}$ to ionize the Universe at high redshifts. \\

\noindent Finally, we note that we performed the same calculations with the luminosity function of $z\sim4$ optically-selected QSOs by~\citet{2011ApJ...728L..26G}. The former parameterization has larger uncertainties than~\citet{2012ApJ...755..169M}, and the resulting non-ionizing emissivity $\epsilon_{UV,24}$ is lower, albeit roughly consistent within the uncertainties, with the values we obtain from Masters. This change of luminosity function does not alter our conclusions.

\section{\lyc~escape fraction from AGN}
\noindent \citet{2015MNRAS.449.4204L} stack $53$ luminous quasars at $z\simeq 2.4$ and apply an average IGM transmission correction. Their corrected spectrum thus represents an intrinsic spectrum of redshift $z\simeq 2.4$ quasars. Quasar spectra appear to be similar in a wide range of redshifts~\citep[][$0.044\leq z\leq4.789$]{2001AJ....122..549V}, so we assume there is no redshift evolution of the quasar continuum from $z\sim2.4$ to our redshift of $z\sim3.1$. The Lusso stack is dominated by non-Xray sources ($43$ out of $53$), while our sample is dominated by Xray AGN ($12$ out of $14$). However, the difference between sources with and without strong Xray emission seems to be in the lines, not in the continuum~\citep[e.g.][]{1998ApJ...498..170G}, and the comparison of our sample to Lusso's is therefore fairly reasonable. We further assume that the average bright quasar has a similar spectrum to the average faint/moderately bright AGN. \\

\begin{table}
\begin{minipage}{84mm}
\centering
\caption{\lyc~escape fraction of the four Type I AGN at redshift $z\sim3.1$ relative to the~\citet{2015MNRAS.449.4204L} IGM corrected intrinsic spectrum. Escape fractions at the restframe effective wavelength ($\lambda^{\lyc}_{eff}$) are stated for a transparent line of sight ($\tau=0$), a ``minimum'' IGM attenuation ($\tau=0.32^{m}$), and a median IGM attenuation ($\tau=0.74^{m}$). A variability correction of $-0.60^{m}$, $-0.62^{m}$ was applied to the \filterr~band flux for AGN5, AGN7. $3\sigma$ upper limits are shown for the non-detections AGN3 and AGN7.}\protect\label{tab:fesc}
\begin{tabular}{ l l c c c c }
\hline
ID & $\lambda^{\lyc}_{eff}$ [\AA] & $f_{esc}^{\tau=0}$ & $f_{esc}^{\tau=0.32}$ & $f_{esc}^{\tau=0.74}$ & $err(f_{esc})$\\
\hline
AGN3 & $867.1$ & $<0.008$ & $<0.011$ & $<0.016$ & \\
AGN5  & $873.5$ & $0.35$ & $0.48$ & $0.73$ & $0.30$\\ 
AGN7  & $873.9$ & $<0.003$ & $<0.004$ & $<0.007$ & \\ 
AGN11 & $879.0$ & $0.15$ & $0.20$ & $0.31$ & $0.27$\\
\hline
\end{tabular}
\end{minipage}
\end{table}

\noindent Under these assumptions we can use the Lusso IGM corrected spectrum as a reasonable approximation of the intrinsic spectrum of AGNs at redshift $\sim3.1$. We can then calculate the escape fraction of \lyc~relative to this intrinsic spectrum, $f_{esc}(\lyc)=f^{\lyc}_{obs}e^{\tau_{IGM}}/f^{\lyc}_{intr}$. We present three resulting fractions, assuming a transparent line of sight to the AGN (no IGM correction), and assuming a ``minimum'' ($\tau=0.32^{m}$) and median ($\tau=0.74^{m}$) IGM attenuation for the observed \filterlyc~(\lyc) flux. The ``minimum'' IGM correction is obtained from the same transmission model as the median~\citep{2008MNRAS.387.1681I,2014MNRAS.442.1805I} as the smallest $0.15\%$ value of the MC simulation. We note that the IGM transmission model used by~\citet[][their Figure 3]{2015MNRAS.449.4204L} on redshift $z=2.4$ objects results in transmission $T_{\lambda}\sim0.62$ at the \lyc~filter's rest-frame wavelengths $\lambda$ of our objects. For the same redshift and wavelength the~\citet{2014MNRAS.442.1805I} IGM model gives a very similar transmission of $T_{\lambda}=0.66$.\\

\noindent The $f_{esc}$ values of all four Type I AGNs at redshift $z\sim3.1$ (AGN3, AGN5, AGN7, AGN11) are shown in Table~\ref{tab:fesc}, with upper limits for AGN3 and AGN7. Here we treat AGN3 as a non-detection in \lyc~due to the unclear nature of the offset emission in the \filterlyc~filter. Since $f^{\lyc}_{obs}$ is obtained from the \filterlyc~band with width $150$\AA, the $f^{\lyc}_{intr}$ is taken as the average over a $150$\AA$/(1+z)$~wavelength bin, centered at the restframe effective wavelength of the \filterlyc~band for each AGN. This is illustrated in Figure~\ref{fig:seds} with gray rectangles at the \filterlyc~position. The indicated errors of $f_{esc}$ in Table~\ref{tab:fesc} are the photometric uncertainty in the \filterlyc~measurement added in quadrature to the average uncertainty in the IGM corrected Lusso spectrum in the relevant wavelength interval. This interval is marked as gray rectangles in Fig.~\ref{fig:seds}. For AGN5 and AGN7 we present $f_{esc}$ from a variability-corrected SED, with formal errors. \\

\noindent The suggested escape fractions for all cases are smaller than unity, consistent with the low emissivities we obtain in Figure~\ref{fig:emissivity}.

\section{Discussion}

\noindent The AGN in our sample fall short of matching ionizing emissivities inferred from assuming average $f_{esc}=1$. They are fainter than high-$z$ quasars however, so it is still possible that bright quasars have escape fractions $f_{esc}=100\%$. If we were to only look at the three brightest AGN in Figure~\ref{fig:colormag}A, which have quasar-like luminosities ($M<-25.0$ in Fig~\ref{fig:colormag}B), the \lyc~detection rate is only $33\%$ ($1$ out of $3$). It is possible that the other two bright AGN (AGN8 at $M_R=-24.99$, and AGN1 at $M_R=-24.35$) are completely obscured by the intervening IGM. Using the brightest AGN as an example (AGN8 at $z\sim3.4$), we calculate the probability to dim a source of a given apparent magnitude $\textit{m}_{\lyc}$ down to the $2\sigma$ limiting magnitude of $\filterlyc(2\sigma)=27.5^{m}$, i.e. below our detection treshold. If we assume that AGN8 will have similar \lyc~brightness as the brightest \lyc~AGN (AGN11, $M_R=-24.48$, $z\sim3.1$), then $\textit{m}_{\lyc}=24.1$. To dim such an object to $\filterlyc(2\sigma)$ we need at least a $3.5$ magnitude obscuration. According to the IGM Monte Carlo simulation from \citet{2008MNRAS.387.1681I} and \citet{2014MNRAS.442.1805I} the probability for full \lyc~obscuration is then $P(\ge3.5)=50.6\%$. We can instead assume that the \lyc~emission is given by the intrinsic~\citet{2015MNRAS.449.4204L} IGM corrected spectrum, which has a $f_{\lyc}/f_{UV}\sim2.0$. Thus, the apparent \lyc~magnitude of this object should be $\textit{m}_{\lyc}=20.3$, and a dimming by $7.3^{m}$ is required, with probability for full obscuration $P(\ge7.3)=40.8\%$. For AGN1 at $z\sim3.8$ the corresponding probabilities are $P(\ge3.5)=80.2\%$ and $P(\ge7.3)=70.7\%$.\\

\noindent All of these probabilities are significant and we cannot reject $f_{esc}=100\%$ for each individual object. To calculate the probability of simultaneous obscuration of two or more objects we assume that the lines of sight and therefore the probabilities are independent. The correlation in the transverse \lya~forest flux in quasar absorption spectra has been measured by e.g.~\citet[][]{2006MNRAS.372.1333D} and~\citet{2006MNRAS.370.1804C}, and is significant at the $3\sigma$ level up to angular separations of $\lesssim6$ comoving Mpc. The minimum angular separation among the five Type I AGNs in our sample which are also non-detections in the \lyc~filter is $221\arcsec$ ($\sim7$ comoving Mpc at $z\sim3.1$). The assumption of independent lines of sight therefore seems reasonable. \\

\noindent The probabilities for both AGN1 and AGN8 to be fully obscured simultaneously are then $P=0.41*0.71\sim29\%$ or $P=0.80*0.71\sim57\%$, which are lower but still significant. For an object at $z\sim3.1$, the probability for full obscuration assuming the Lusso spectrum is only $P(\ge7.3)=10.3\%$. The probability to fully obscure the two Type I AGNs which are members of the proto-cluster and show non-detections in \lyc~(AGN3, AGN7, $z\sim3.1$) is then $P(\ge7.3)=0.1^2=1.1\%$, which seems very low. To obscure all five Type I AGNs (AGN1, AGN3, AGN7, AGN8, AGN13, $z\sim3.1$-$3.46$) the probability is $P(\ge7.3)=0.71*0.10*0.41*0.10*0.49\sim0.1\%$, which is negligible. It therefore seems unlikely that IGM obscuration could account for the low detection rate we observe, suggesting that the average escape fraction from moderately bright/faint AGN is not unity.\\

\noindent One should further consider that there is a non-zero probability to have at least two foreground contaminants in our sample. If the detections we observe in the \filterlyc~filter come from low-$z$ galaxies then this would only strengthen our conclusion that the observed detection rates are not easily explained with IGM obscuration, and therefore the data seem inconsistent with $\left<f_{esc}\right>=1$.\\

\noindent Recently,~\citet{2016arXiv160309351C} stacked $1669$ bright quasars from the SDSS BOSS survey and find an average $f_{esc}=0.7$. Their sample is however at a higher redshift and a brighter magnitude range. Their faintest quasars are $\sim2^{m}$ brighter than our brightest AGNs.

\section{Conclusions}
\noindent We present Subaru/SuprimeCam \filterb\filterv\filterr\filteri\filterz,\lya,\lyc~broadband and narrowband photometry for a sample of $14$ spectroscopically confirmed $z>3.08$ AGN from the SSA22 field, as well as six spectra to complement the available literature. Four AGN are detected in the \lyc~filter. Two \lyc~detections show offsets and are likely stellar in nature. For the remaining two AGN the \lyc~is spatially coincident with the UV continuum and \lya~emission, with no measureable offsets between UV and \lya~to \lyc~within our astrometry uncertainty. The statistical probability to have at least three foreground contaminants in this sample is $0.6\%$. \\

\noindent If the nature of the \lyc~detection in the two AGNs showing a spatial offset between \lyc~and UV is indeed stellar (probability for contamination $P(\ge2)=5.5\%$), then the detection rate of stellar \lyc~from AGN seems fairly high - at least $\sim0.07^{+0.24}_{-0.01}$ $(1/14)$ or even $\sim0.14^{+0.33}_{-0.05}$ $(2/14)$. This is comparable to star forming LBGs and LAEs at the same redshift. The detection rate among Type I AGNs is $0.29^{+0.66}_{-0.10}$ $(2/7)$, which is higher than star forming galaxies, although not significantly so within the uncertanties. \\

\noindent Through stacking we obtain average flux density ratios and ionizing emissivity for e.g. \lyc, Type I, Xray, and $z=3.1$ subgroups, statistically corrected for foreground contamination. The largest emissivity we obtain accounts for at most $12\%$ and as little as $5\%$ of the ionization budget predicted by Ly$\alpha$ forest observations in quasar spectra at the same redshift. Our direct measurement of the ionizing emissivity of Type I AGNs is on average a factor of $\sim7$ lower than the emissivity obtained from assuming a double power law with $f_{esc}=1$. Our sample indicates that the average $f_{esc}$ does not seem to be unity. It is possible for individual AGNs to be fully obscured by the intervening IGM, however the probability that e.g. all five out of the seven Type I AGNs in the sample are fully obscured is negligible, $P=0.1\%$. This may suggest that faint AGN, even if more numerous than previously thought, are not sufficient to completely dominate the ionization budget of the Universe at high redshifts. \\

\noindent Examining the properties of the AGN we find that three are significantly variable and $12$ are Xray sources. The small numbers prevent us from discerning any possible connection between variability and Xray emission to \lyc~escape.

\section*{Acknowledgments}
GM and II are supported by JSPS KAKENHI Grant number: 24244018. GM acknowledges support by the Swedish Research Council (Vetenskapsr\aa det). AKI is supported by JSPS KAKENHI Grant number: 26287034.

\noindent A part of this research has made use of the NASA/ IPAC Infrared Science Archive, which is operated by the Jet Propulsion Laboratory, California Institute of Technology, under contract with NASA.

\noindent We extend a warm thank you to Yuichi Matsuda, Katsuki Kousai, Elisabeta Lusso, Cristian Saez, and Alice Shapley for providing us with their spectra, and with some clarifications on their work and helpful advice. We also thank Andrea Grazian for useful suggestions and comments on this work.

\appendix
\section{Spectra of the AGN sample}
\noindent The majority of the spectra of the presented sample can be found in the catalogs of~\citet{2015MNRAS.450.2615S},~\citet{2012MNRAS.425..878M},~\citet{2005Natur.436..227W},~\citet{2013ApJ...765...47N} or the VVDS online database. The missing spectra are for AGN1, AGN2, AGN3, AGN4, and AGN5 which we show here for completion. 

\begin{figure}
\centering\tiny
\includegraphics[width=85mm]{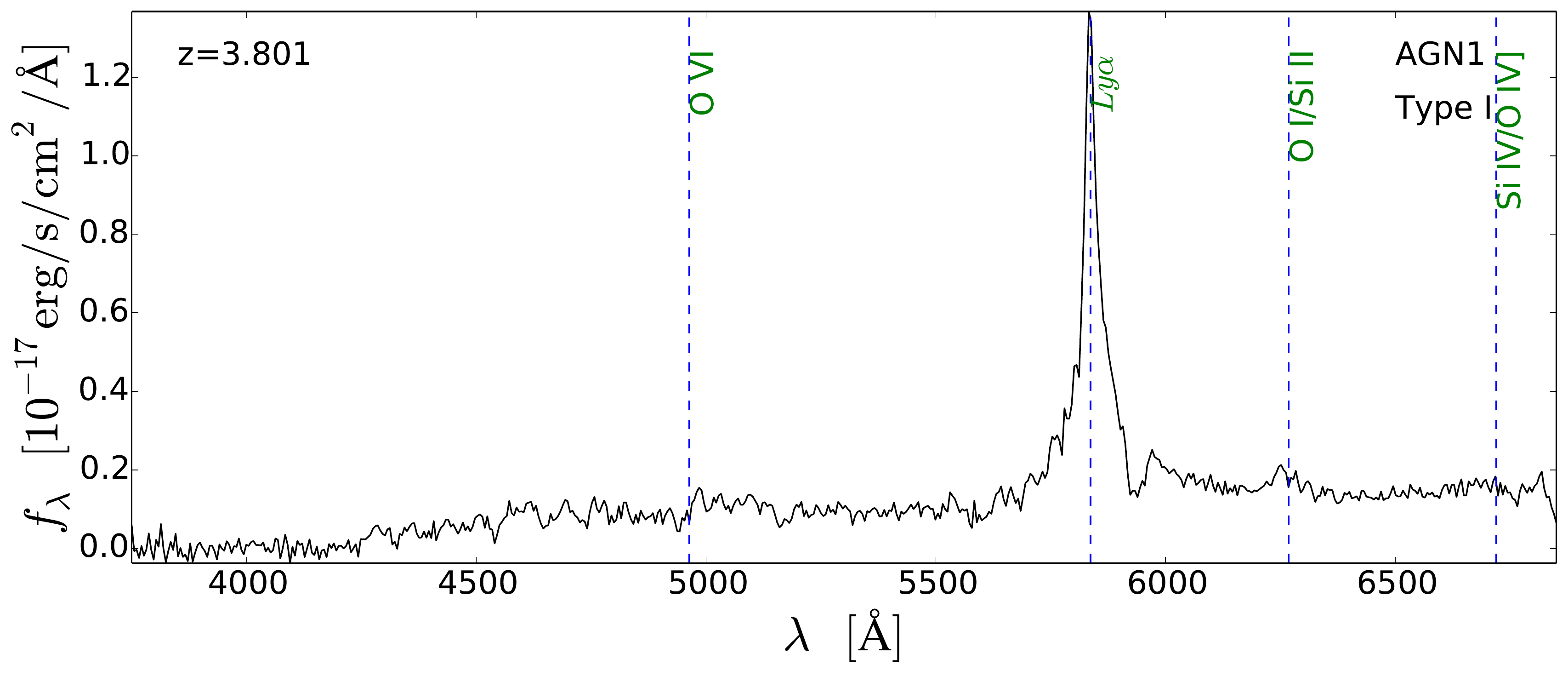}
\includegraphics[width=85mm]{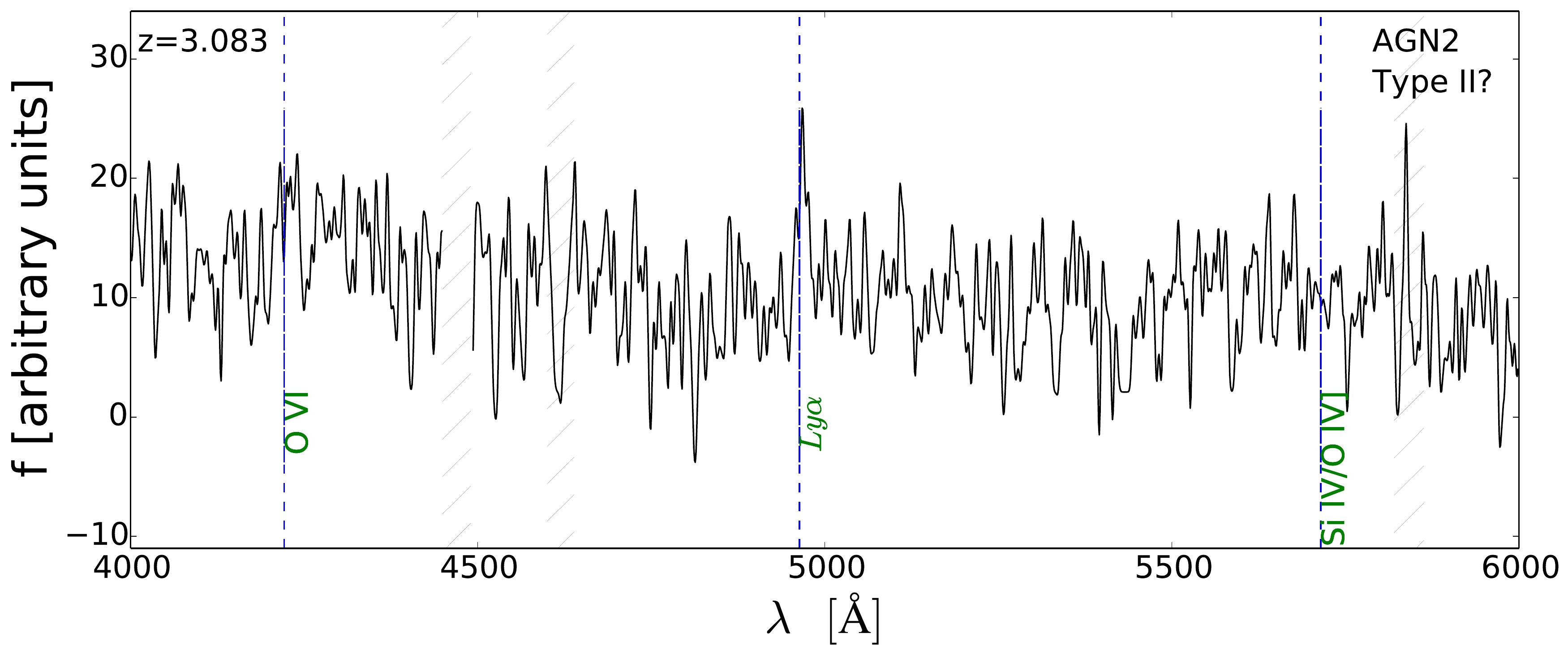}
\includegraphics[width=85mm]{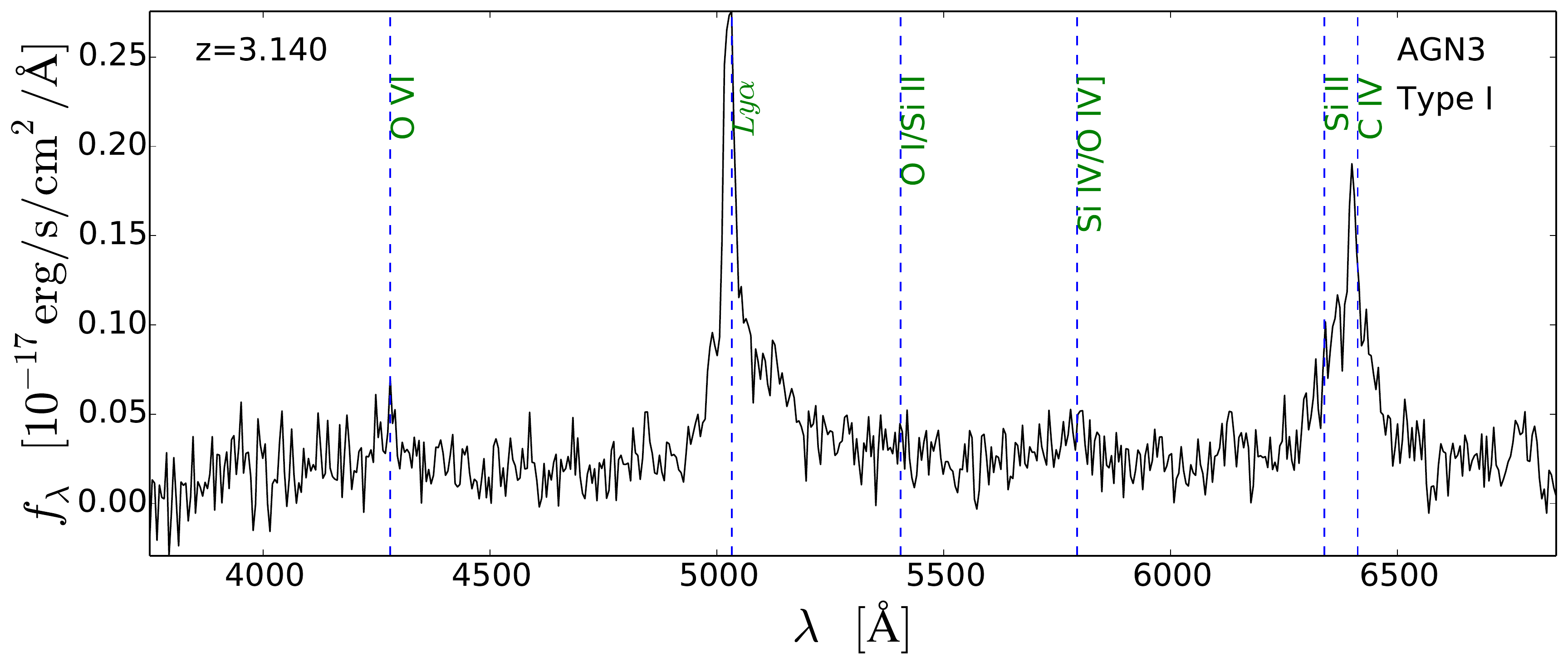}
\includegraphics[width=85mm]{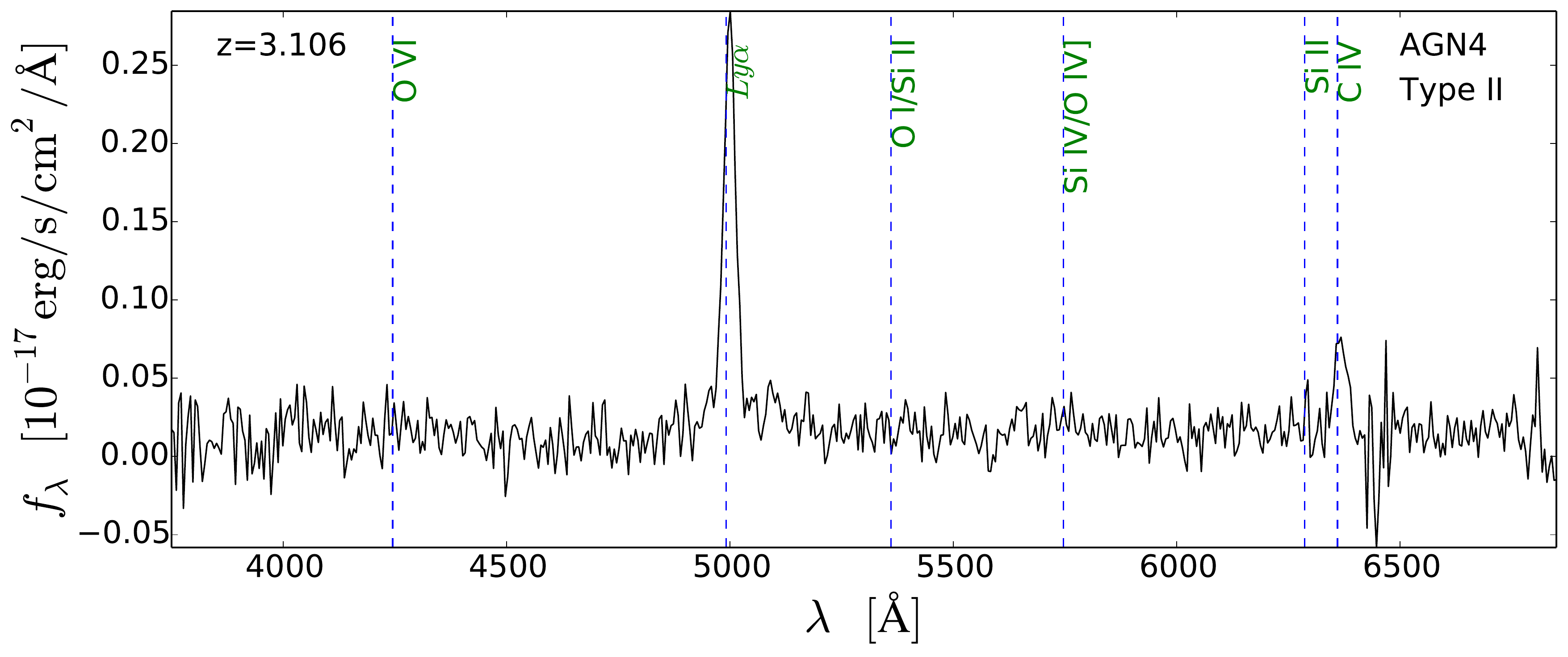}
\includegraphics[width=85mm]{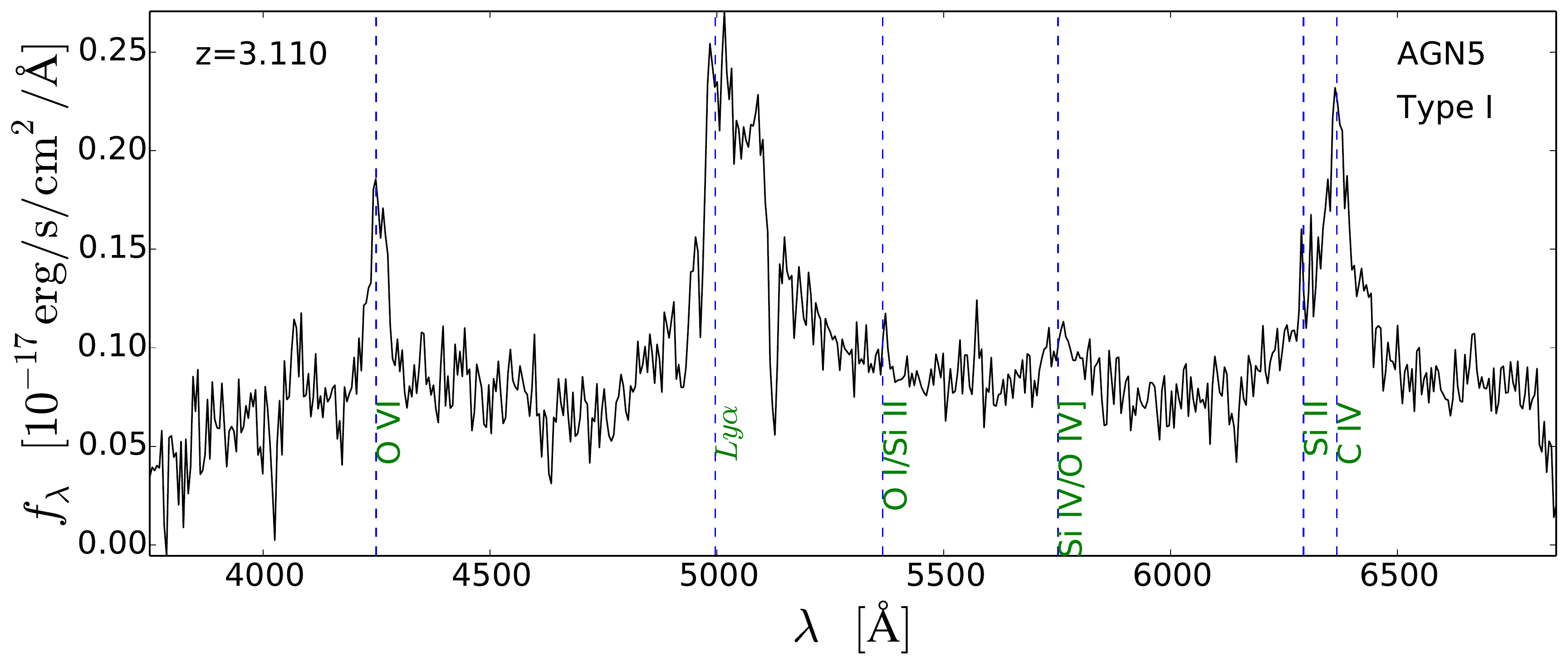}
\caption{Complementary spectra of the AGN sample. Data are from DEIMOS 2004 (AGN2), VIMOS 2006 (AGN1, AGN4), VIMOS 2008 (AGN3, AGN5). These spectra are used solely to demonstrate the redshift and classification of these AGN.}\protect\label{fig:spectra}
\end{figure}

\section{Mosaics of non-detections in \lyc}
\noindent Here we show the complementary images of the \lyc~non-detections. For high-$z$ AGN the \filterlya$-$\filterbv images do not trace continuum subtracted \lya, so we show instead the \filterlya~images which simply sample the UV continuum. 

\begin{figure}
\centering\tiny
\includegraphics[width=85mm]{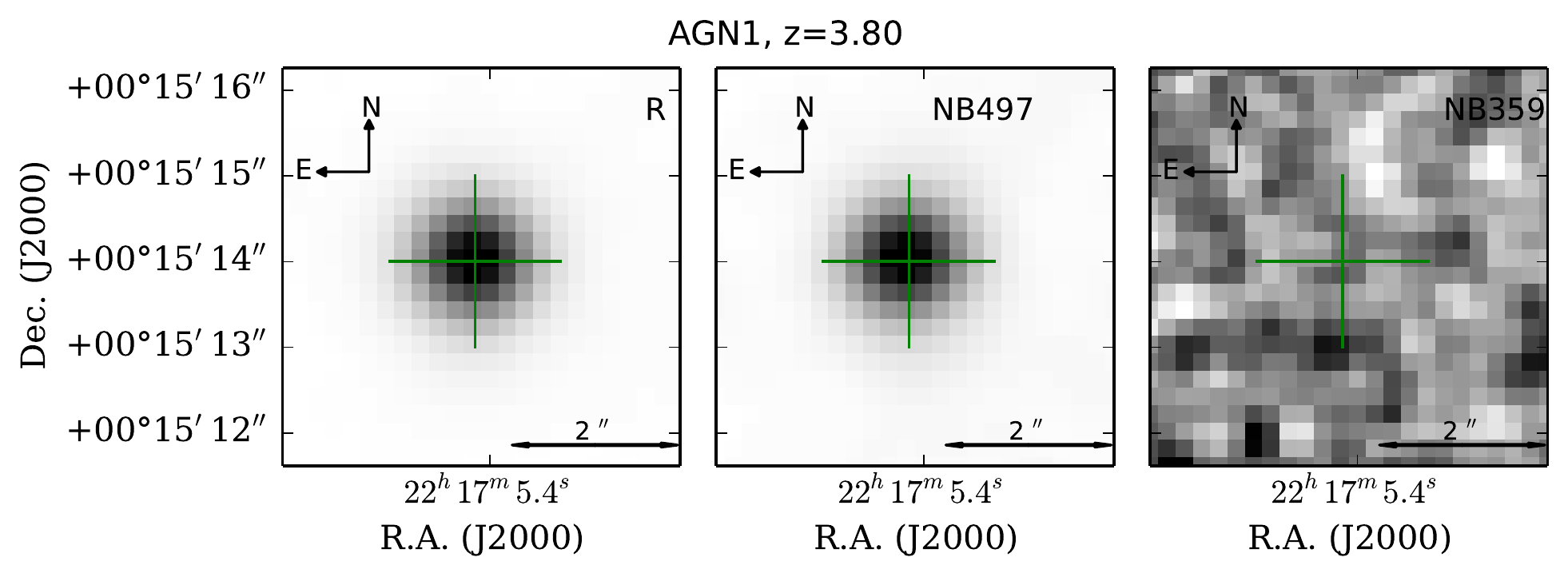}
\includegraphics[width=85mm]{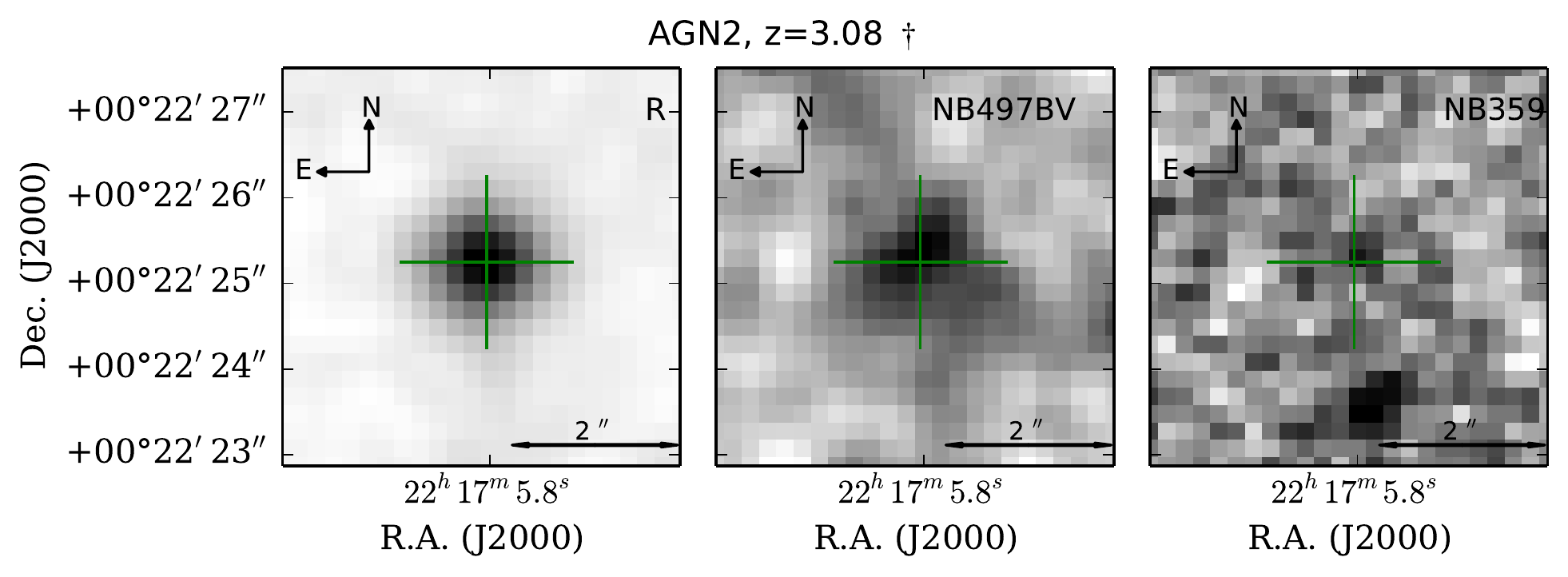}
\includegraphics[width=85mm]{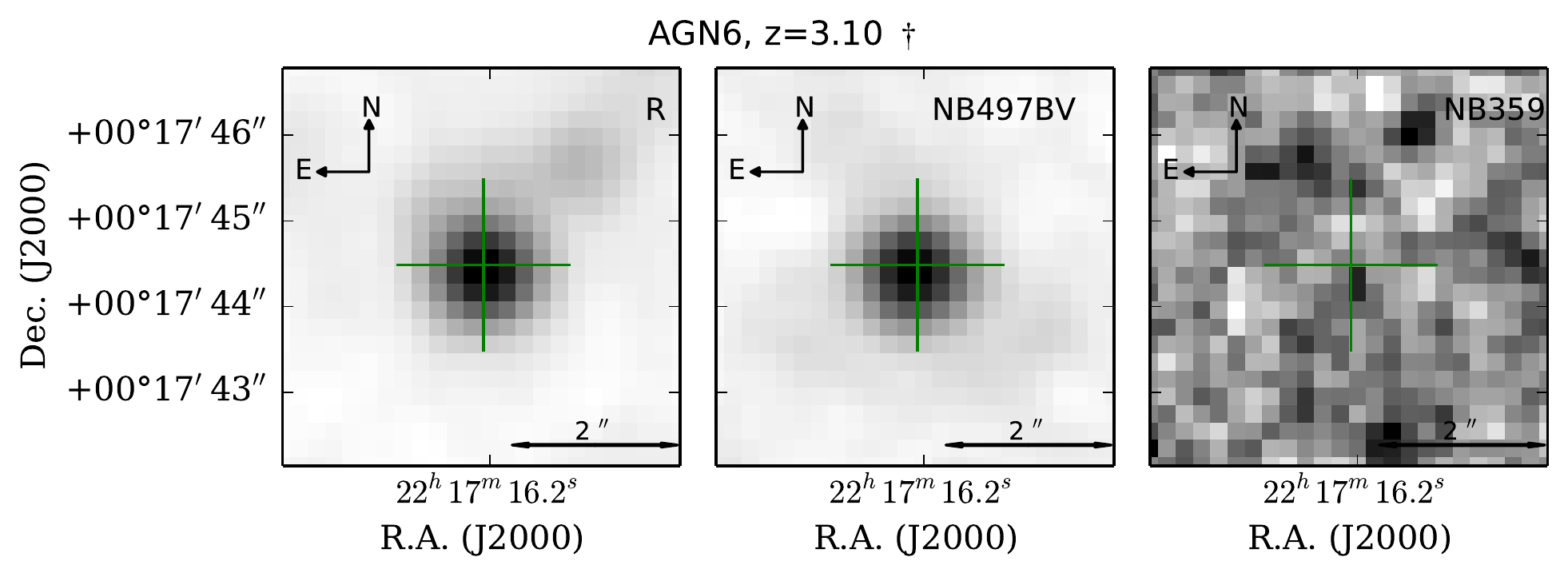}
\includegraphics[width=85mm]{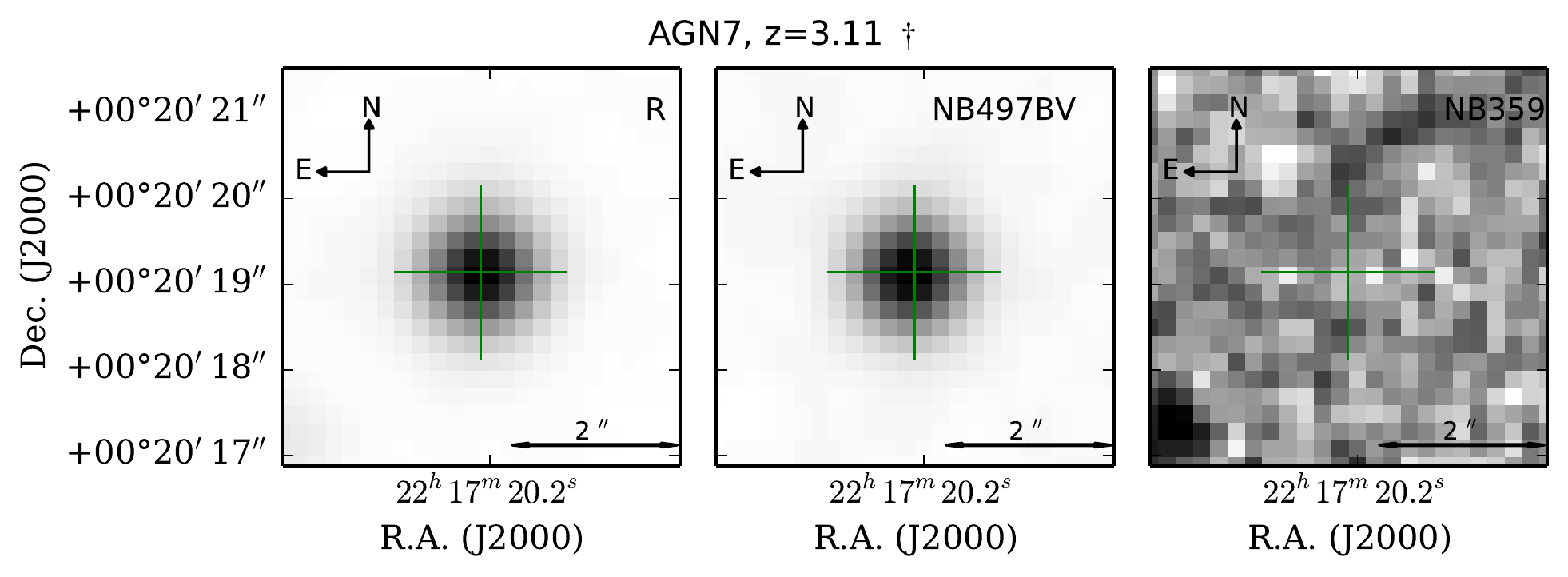}
\includegraphics[width=85mm]{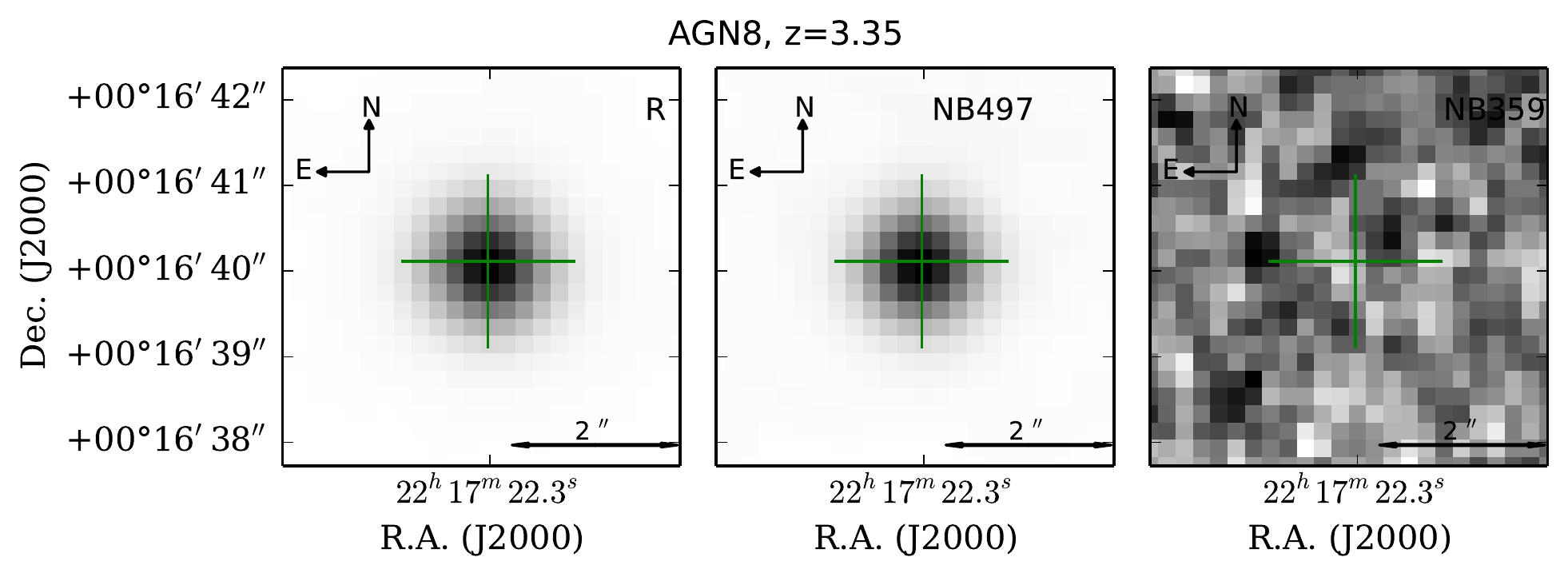}
\includegraphics[width=85mm]{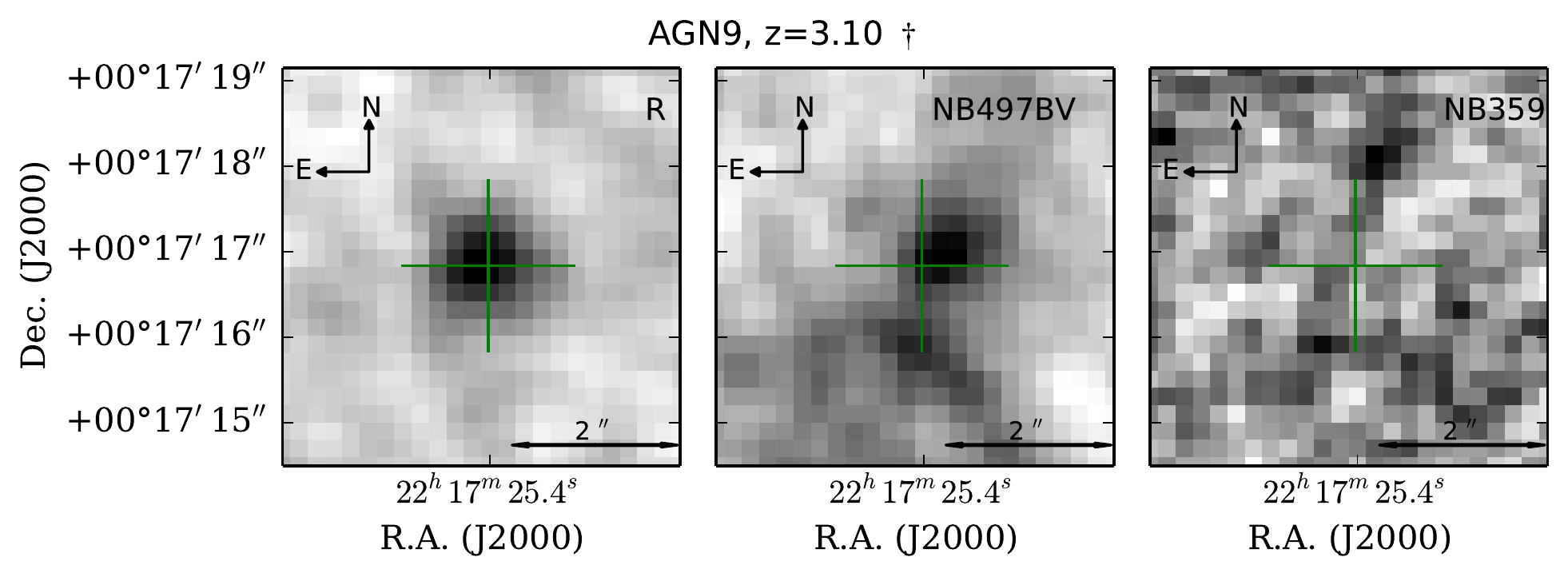}
\caption{Complementary images for all non-detections in \lyc. The \filterlya$-$\filterbv~image (middle column) shows continuum subtracted \lya~emission for redshifts $3.06\leq z\leq3.12$, indicated with a dagger. For higher-$z$ AGN we show the \filterlya~image instead. The \filterlyc~image (right column) shows \lyc. The contrast and intensity settings are the same as in Figure~\ref{fig:contours}.}\protect\label{fig:mosaic_non_detect}
\end{figure}

\begin{figure}
\centering\tiny
\includegraphics[width=85mm]{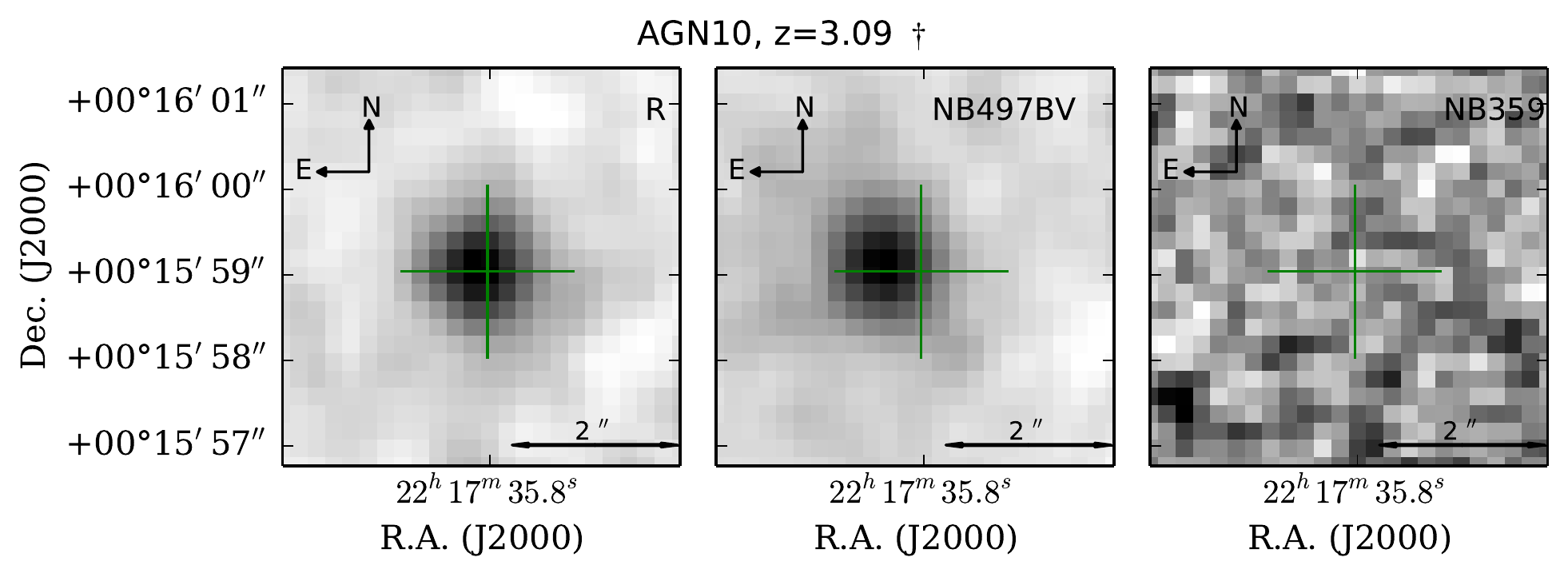}
\includegraphics[width=85mm]{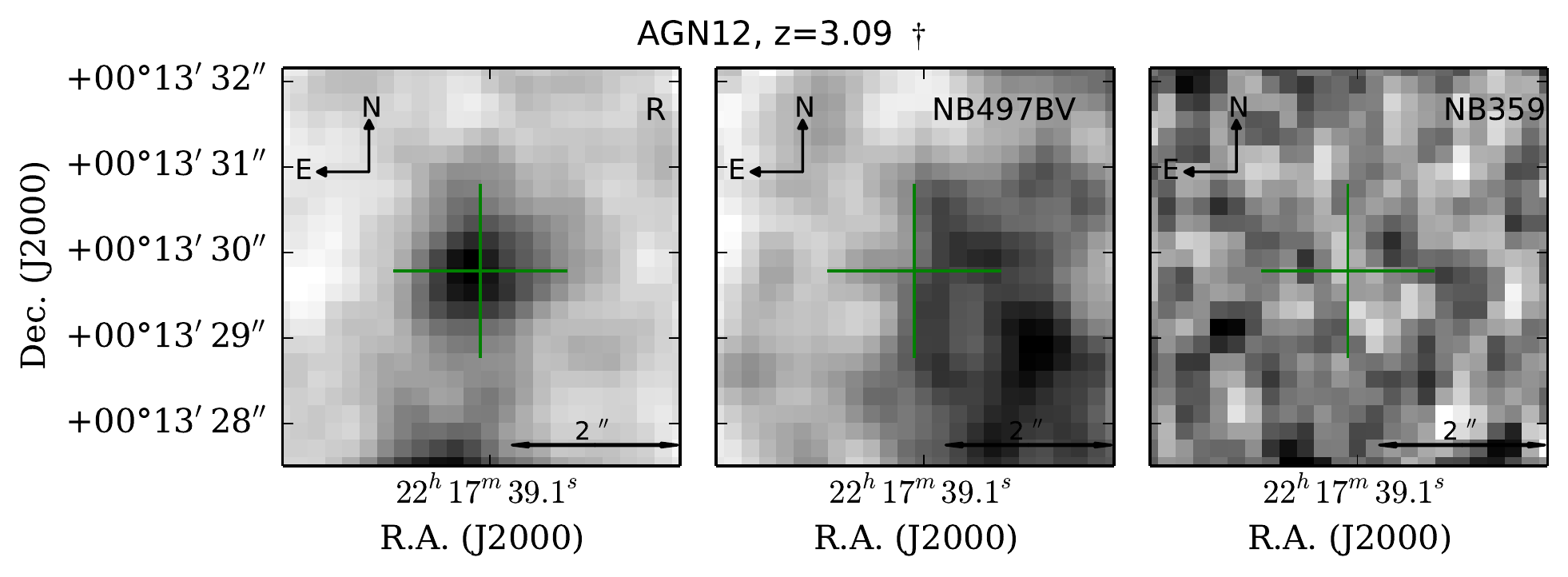}
\includegraphics[width=85mm]{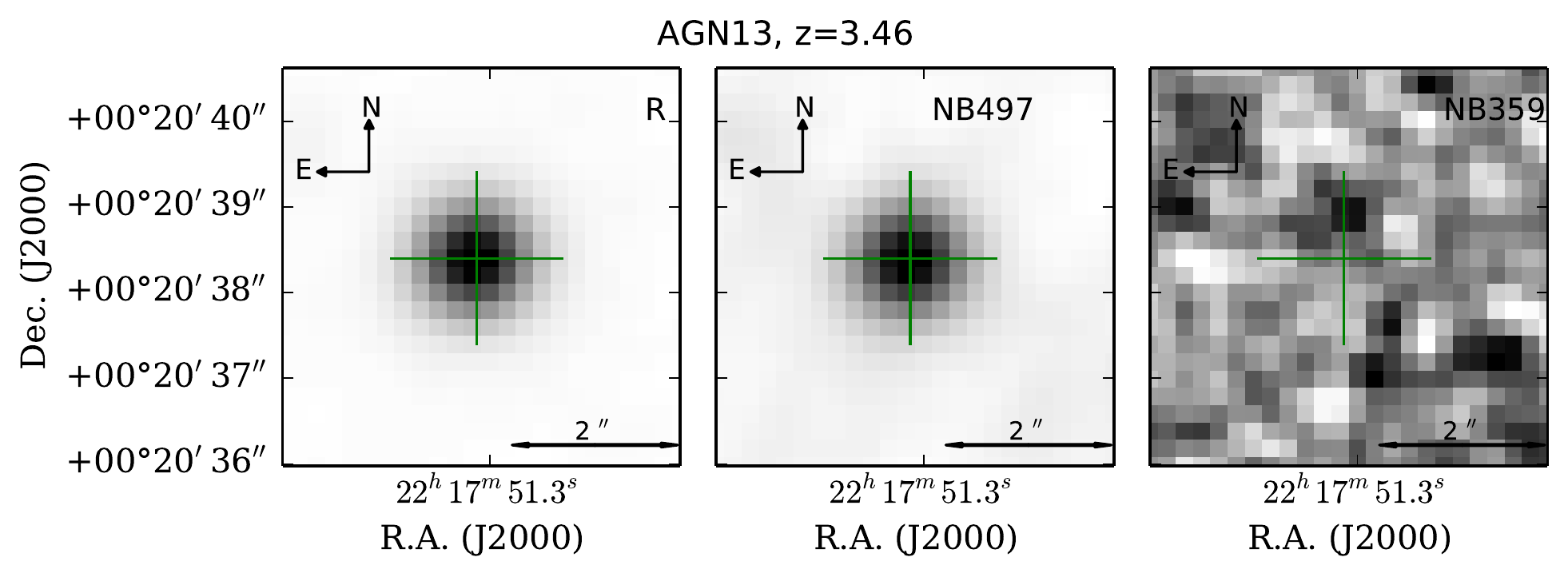}
\includegraphics[width=85mm]{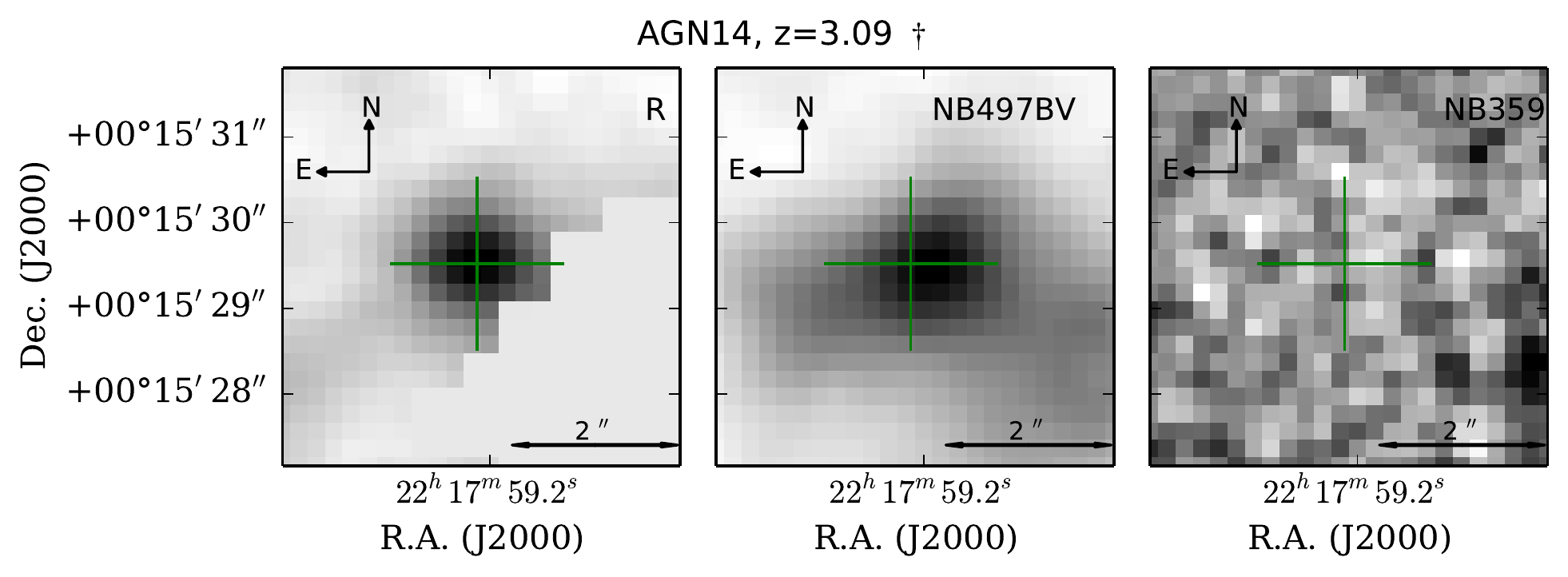}
\contcaption{}
\end{figure}

\section{Extra images of AGN3 and AGN4}
\noindent To investigate the nature of the offset \lyc~detection in these AGNs we present a selection of additional images.
\begin{figure}
\centering\tiny
\includegraphics[width=85mm]{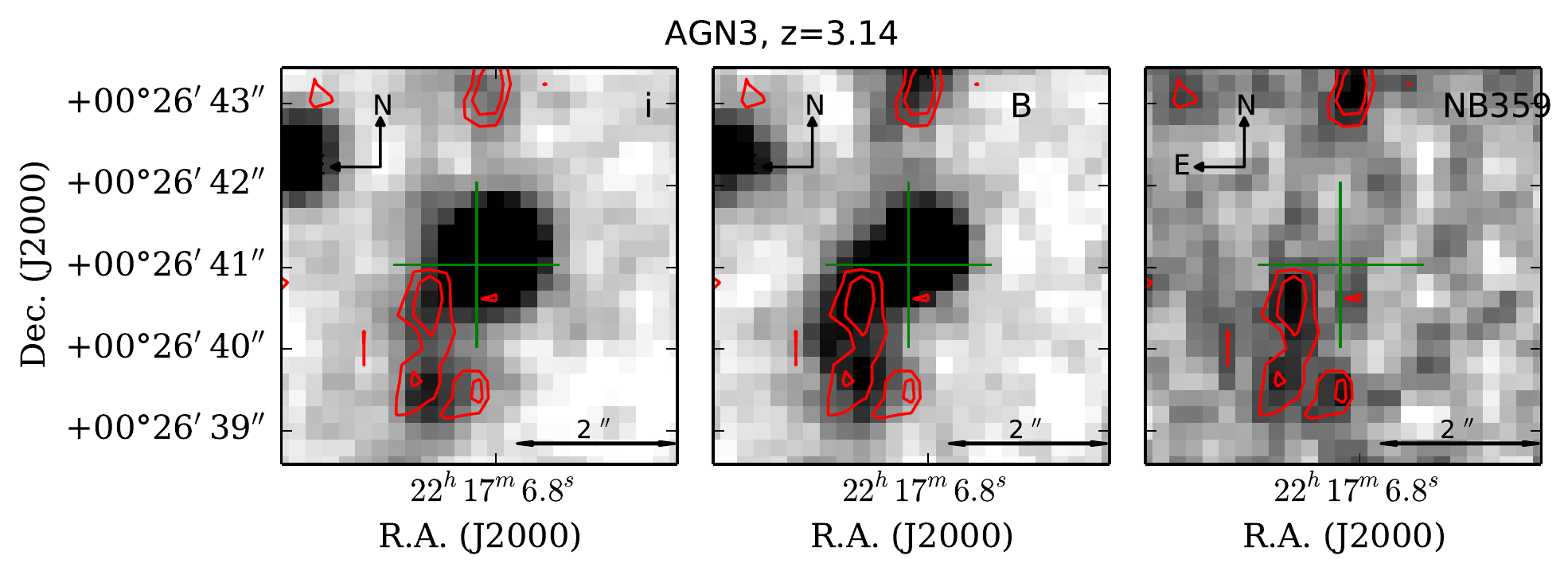}
\includegraphics[width=85mm]{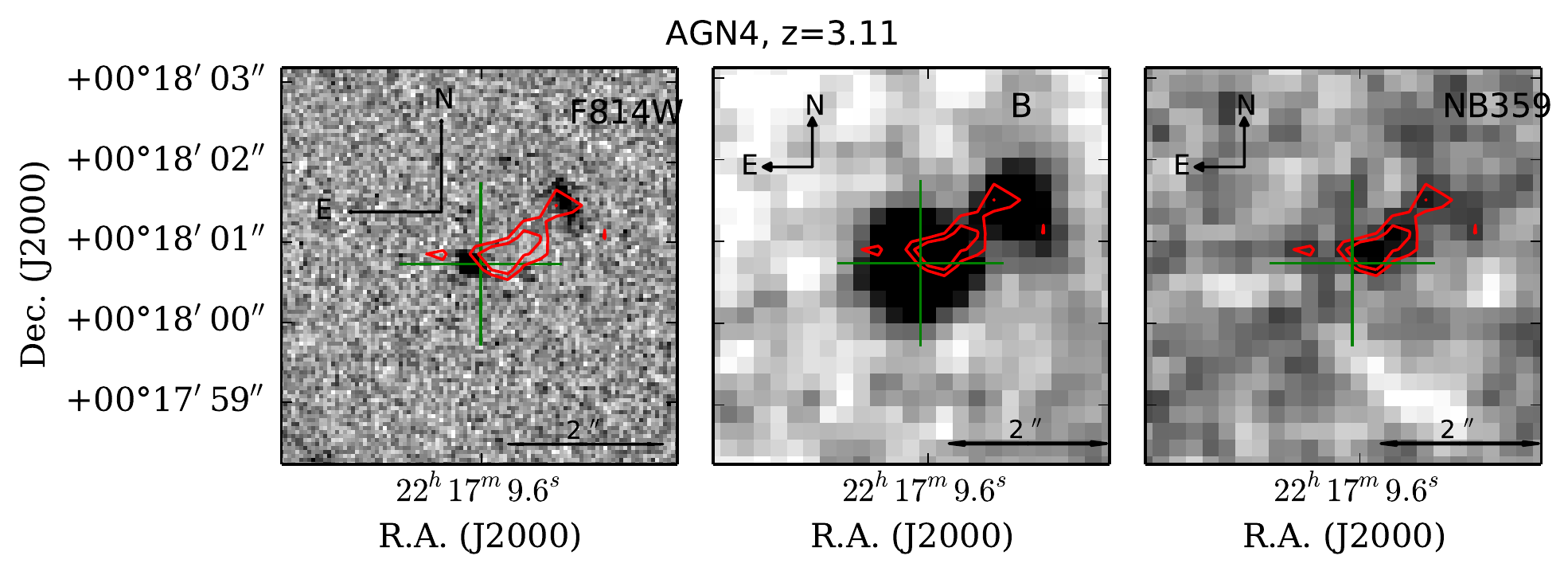}
\caption{Additional images for stellar \lyc~AGN3 (\filteri,\filterb) and \lyc~AGN4 (HST $F814W$, \filterb).}\protect\label{fig:additional}
\end{figure}
\section{SEDs of LABs AGN10 and AGN14}
\begin{figure}
\centering\tiny
\includegraphics[width=85mm]{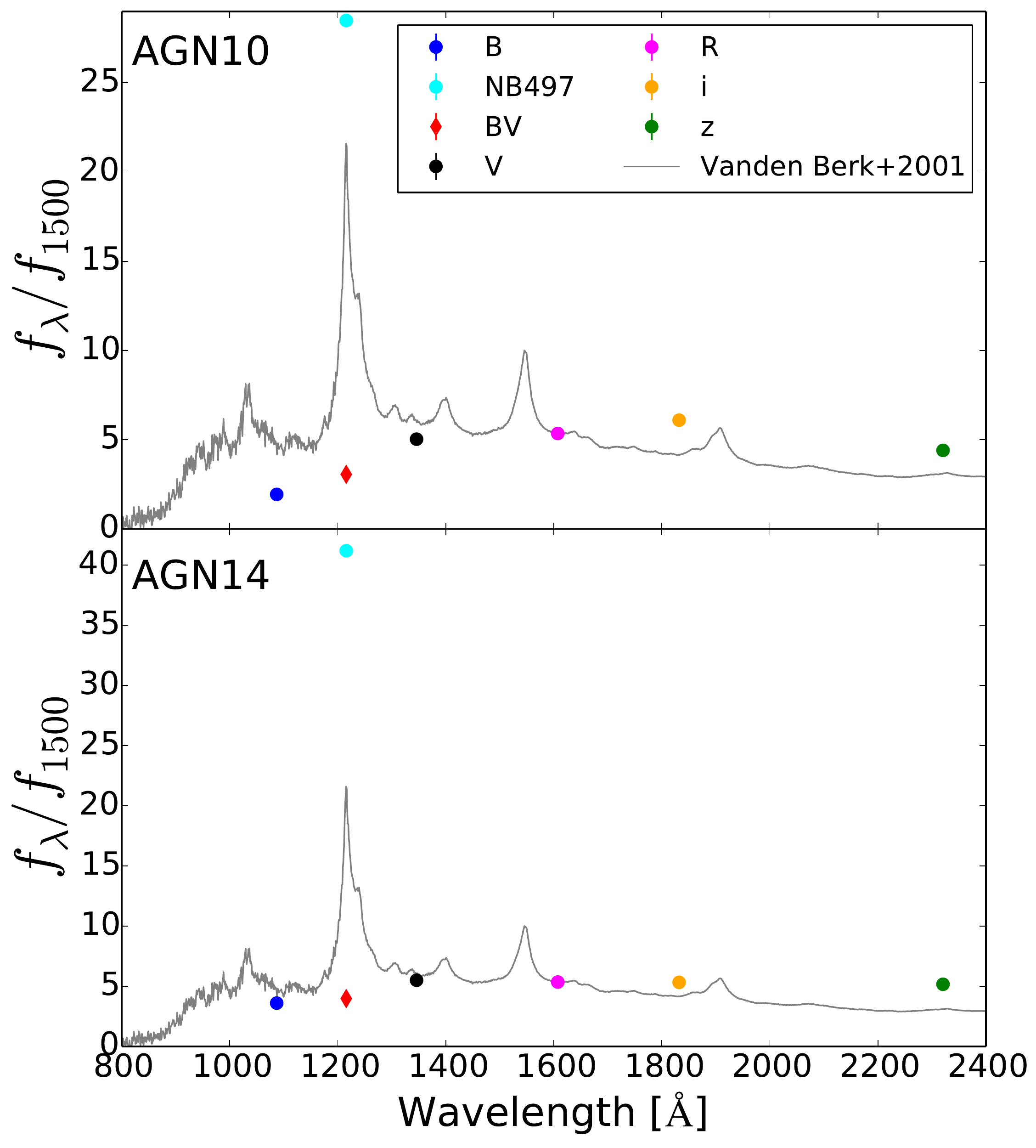}
\caption{SEDs of LABs AGN10 and AGN14 showing their red spectrum and extreme \lya.}\protect\label{fig:hostdominated}
\end{figure}

\noindent These two AGNs are also Lyman $\alpha$ Blobs and show extreme observed \ew~and extended \lya~emission. Their SEDs in Figure~\ref{fig:hostdominated} reveal a spectrum much redder than the typical quasar spectrum.

\section{Photometry catalog of the AGN sample}
\noindent This work comes with an online catalog of \filterb\filterv\filterr\filteri\filterz~broadband photometry, narrowband \filterlyc~(\lyc~at $z>3.06$) and \filterlya~(\lya~at $3.06<z<3.127$) for all $14$ AGN in our sample. The catalog contains AB magnitude measurements from asymptotic growth curve measurements (total mag) and from a $\diameter=1.2\arcsec$ fixed aperture photometry. 

\bibliographystyle{mn2e.bst}
\bibliography{agn}

\begin{thebibliography}{}

\bibitem[\protect\citeauthoryear{{Alvarez}, {Finlator} \& {Trenti}}{{Alvarez}
  et~al.}{2012}]{2012ApJ...759L..38A}
{Alvarez} M.~A.,  {Finlator} K.,    {Trenti} M.,  2012, \apjl, 759, L38

\bibitem[\protect\citeauthoryear{{Becker} \& {Bolton}}{{Becker} \&
  {Bolton}}{2013}]{2013MNRAS.436.1023B}
{Becker} G.~D.,  {Bolton} J.~S.,  2013, \mnras, 436, 1023

\bibitem[\protect\citeauthoryear{{Becker}, {Fan}, {White} \& et. al.}{{Becker}
  et~al.}{2001}]{2001AJ....122.2850B}
{Becker} R.~H.,  {Fan} X.,  {White} R.~L.,    et. al. 2001, \aj, 122, 2850

\bibitem[\protect\citeauthoryear{{Bongiorno}, {Zamorani}, {Gavignaud} \& et.
  al.}{{Bongiorno} et~al.}{2007}]{2007A&A...472..443B}
{Bongiorno} A.,  {Zamorani} G.,  {Gavignaud} I.,    et. al. 2007, \aap, 472,
  443

\bibitem[\protect\citeauthoryear{{Bouwens}, {Illingworth}, {Oesch} \& et.
  al.}{{Bouwens} et~al.}{2015}]{2015ApJ...811..140B}
{Bouwens} R.~J.,  {Illingworth} G.~D.,  {Oesch} P.~A.,    et. al. 2015, \apj,
  811, 140

\bibitem[\protect\citeauthoryear{{Coppolani}, {Petitjean}, {Stoehr},
  {Rollinde}, {Pichon}, {Colombi}, {Haehnelt}, {Carswell} \&
  {Teyssier}}{{Coppolani} et~al.}{2006}]{2006MNRAS.370.1804C}
{Coppolani} F.,  {Petitjean} P.,  {Stoehr} F.,  {Rollinde} E.,  {Pichon} C.,
  {Colombi} S.,  {Haehnelt} M.~G.,  {Carswell} B.,    {Teyssier} R.,  2006,
  \mnras, 370, 1804

\bibitem[\protect\citeauthoryear{{Cowie}, {Barger} \& {Trouille}}{{Cowie}
  et~al.}{2009}]{2009ApJ...692.1476C}
{Cowie} L.~L.,  {Barger} A.~J.,    {Trouille} L.,  2009, \apj, 692, 1476

\bibitem[\protect\citeauthoryear{{Cristiani}, {Serrano}, {Fontanot},
  {Koothrappali}, {Vanzella} \& {Monaco}}{{Cristiani}
  et~al.}{2016}]{2016arXiv160309351C}
{Cristiani} S.,  {Serrano} L.~M.,  {Fontanot} F.,  {Koothrappali} R.~R.,
  {Vanzella} E.,    {Monaco} P.,  2016, ArXiv e-prints

\bibitem[\protect\citeauthoryear{{De Cicco}, {Paolillo}, {Covone} \& et
  al.}{{De Cicco} et~al.}{2015}]{2015A&A...574A.112D}
{De Cicco} D.,  {Paolillo} M.,  {Covone} G.,    et al. 2015, \aap, 574, A112

\bibitem[\protect\citeauthoryear{{de Vries}, {Becker} \& {White}}{{de Vries}
  et~al.}{2003}]{2003AJ....126.1217D}
{de Vries} W.~H.,  {Becker} R.~H.,    {White} R.~L.,  2003, \aj, 126, 1217

\bibitem[\protect\citeauthoryear{{de Vries}, {Becker}, {White} \& et al.}{{de
  Vries} et~al.}{2005}]{2005AJ....129..615D}
{de Vries} W.~H.,  {Becker} R.~H.,  {White} R.~L.,    et al. 2005, \aj, 129,
  615

\bibitem[\protect\citeauthoryear{{D'Odorico}, {Viel}, {Saitta}, {Cristiani},
  {Bianchi}, {Boyle}, {Lopez}, {Maza} \& {Outram}}{{D'Odorico}
  et~al.}{2006}]{2006MNRAS.372.1333D}
{D'Odorico} V.,  {Viel} M.,  {Saitta} F.,  {Cristiani} S.,  {Bianchi} S.,
  {Boyle} B.,  {Lopez} S.,  {Maza} J.,    {Outram} P.,  2006, \mnras, 372, 1333

\bibitem[\protect\citeauthoryear{{Duncan} \& {Conselice}}{{Duncan} \&
  {Conselice}}{2015}]{2015MNRAS.451.2030D}
{Duncan} K.,  {Conselice} C.~J.,  2015, \mnras, 451, 2030

\bibitem[\protect\citeauthoryear{{Fan}, {Narayanan}, {Lupton}, {Strauss} \& et.
  al.}{{Fan} et~al.}{2001}]{2001AJ....122.2833F}
{Fan} X.,  {Narayanan} V.~K.,  {Lupton} R.~H.,  {Strauss} M.~A.,    et. al.
  2001, \aj, 122, 2833

\bibitem[\protect\citeauthoryear{{Faucher-Gigu{\`e}re}, {Lidz}, {Zaldarriaga}
  \& et. al.}{{Faucher-Gigu{\`e}re} et~al.}{2009}]{2009ApJ...703.1416F}
{Faucher-Gigu{\`e}re} C.-A.,  {Lidz} A.,  {Zaldarriaga} M.,    et. al. 2009,
  \apj, 703, 1416

\bibitem[\protect\citeauthoryear{{Favre}, {Courvoisier} \& {Paltani}}{{Favre}
  et~al.}{2005}]{2005A&A...443..451F}
{Favre} P.,  {Courvoisier} T.~J.-L.,    {Paltani} S.,  2005, \aap, 443, 451

\bibitem[\protect\citeauthoryear{{Fontanot}, {Cristiani}, {Pfrommer} \& et.
  al.}{{Fontanot} et~al.}{2014}]{2014MNRAS.438.2097F}
{Fontanot} F.,  {Cristiani} S.,  {Pfrommer} C.,    et. al. 2014, \mnras, 438,
  2097

\bibitem[\protect\citeauthoryear{{Fontanot}, {Cristiani} \&
  {Vanzella}}{{Fontanot} et~al.}{2012}]{2012MNRAS.425.1413F}
{Fontanot} F.,  {Cristiani} S.,    {Vanzella} E.,  2012, \mnras, 425, 1413

\bibitem[\protect\citeauthoryear{{Gabel}, {Kraemer}, {Crenshaw} \& et
  al.}{{Gabel} et~al.}{2005}]{2005ApJ...631..741G}
{Gabel} J.~R.,  {Kraemer} S.~B.,  {Crenshaw} D.~M.,    et al. 2005, \apj, 631,
  741

\bibitem[\protect\citeauthoryear{{Garc{\'{\i}}a-Gonz{\'a}lez},
  {Alonso-Herrero}, {P{\'e}rez-Gonz{\'a}lez} \& et
  al.}{{Garc{\'{\i}}a-Gonz{\'a}lez} et~al.}{2015}]{2015MNRAS.446.3199G}
{Garc{\'{\i}}a-Gonz{\'a}lez} J.,  {Alonso-Herrero} A.,
  {P{\'e}rez-Gonz{\'a}lez} P.~G.,    et al. 2015, \mnras, 446, 3199

\bibitem[\protect\citeauthoryear{{Gavignaud}, {Wisotzki}, {Bongiorno} \& et
  al.}{{Gavignaud} et~al.}{2008}]{2008A&A...492..637G}
{Gavignaud} I.,  {Wisotzki} L.,  {Bongiorno} A.,    et al. 2008, \aap, 492, 637

\bibitem[\protect\citeauthoryear{{Geach}, {Matsuda}, {Smail} \& et.
  al.}{{Geach} et~al.}{2005}]{2005MNRAS.363.1398G}
{Geach} J.~E.,  {Matsuda} Y.,  {Smail} I.,    et. al. 2005, \mnras, 363, 1398

\bibitem[\protect\citeauthoryear{{Gehrels}}{{Gehrels}}{1986}]{1986ApJ...303..336G}
{Gehrels} N.,  1986, \apj, 303, 336

\bibitem[\protect\citeauthoryear{{Giallongo}, {Grazian}, {Fiore} \& et.
  al.}{{Giallongo} et~al.}{2015}]{2015A&A...578A..83G}
{Giallongo} E.,  {Grazian} A.,  {Fiore} F.,    et. al. 2015, \aap, 578, A83

\bibitem[\protect\citeauthoryear{{Giallongo}, {Menci}, {Fiore} \& et.
  al.}{{Giallongo} et~al.}{2012}]{2012ApJ...755..124G}
{Giallongo} E.,  {Menci} N.,  {Fiore} F.,    et. al. 2012, \apj, 755, 124

\bibitem[\protect\citeauthoryear{{Glikman}, {Djorgovski}, {Stern} \& et.
  al.}{{Glikman} et~al.}{2011}]{2011ApJ...728L..26G}
{Glikman} E.,  {Djorgovski} S.~G.,  {Stern} D.,    et. al. 2011, \apjl, 728,
  L26

\bibitem[\protect\citeauthoryear{{Grazian}, {Giallongo}, {Gerbasi} \& et.
  al.}{{Grazian} et~al.}{2015}]{2015arXiv150901101G}
{Grazian} A.,  {Giallongo} E.,  {Gerbasi} R.,    et. al. 2015, ArXiv e-prints

\bibitem[\protect\citeauthoryear{{Green}}{{Green}}{1998}]{1998ApJ...498..170G}
{Green} P.~J.,  1998, \apj, 498, 170

\bibitem[\protect\citeauthoryear{{Haardt} \& {Madau}}{{Haardt} \&
  {Madau}}{1996}]{1996ApJ...461...20H}
{Haardt} F.,  {Madau} P.,  1996, \apj, 461, 20

\bibitem[\protect\citeauthoryear{{Inoue} \& {Iwata}}{{Inoue} \&
  {Iwata}}{2008}]{2008MNRAS.387.1681I}
{Inoue} A.~K.,  {Iwata} I.,  2008, \mnras, 387, 1681

\bibitem[\protect\citeauthoryear{{Inoue}, {Iwata} \& {Deharveng}}{{Inoue}
  et~al.}{2006}]{2006MNRAS.371L...1I}
{Inoue} A.~K.,  {Iwata} I.,    {Deharveng} J.-M.,  2006, \mnras, 371, L1

\bibitem[\protect\citeauthoryear{{Inoue}, {Kousai}, {Iwata}, {Matsuda} \& et.
  al.}{{Inoue} et~al.}{2011}]{2011MNRAS.411.2336I}
{Inoue} A.~K.,  {Kousai} K.,  {Iwata} I.,  {Matsuda} Y.,    et. al. 2011,
  \mnras, 411, 2336

\bibitem[\protect\citeauthoryear{{Inoue}, {Shimizu}, {Iwata} \&
  {Tanaka}}{{Inoue} et~al.}{2014}]{2014MNRAS.442.1805I}
{Inoue} A.~K.,  {Shimizu} I.,  {Iwata} I.,    {Tanaka} M.,  2014, \mnras, 442,
  1805

\bibitem[\protect\citeauthoryear{{Khaire} \& {Srianand}}{{Khaire} \&
  {Srianand}}{2015}]{2015MNRAS.451L..30K}
{Khaire} V.,  {Srianand} R.,  2015, \mnras, 451, L30

\bibitem[\protect\citeauthoryear{{Klesman} \& {Sarajedini}}{{Klesman} \&
  {Sarajedini}}{2007}]{2007ApJ...665..225K}
{Klesman} A.,  {Sarajedini} V.,  2007, \apj, 665, 225

\bibitem[\protect\citeauthoryear{{Lehmer}, {Alexander}, {Chapman} \& et.
  al.}{{Lehmer} et~al.}{2009}]{2009MNRAS.400..299L}
{Lehmer} B.~D.,  {Alexander} D.~M.,  {Chapman} S.~C.,    et. al. 2009, \mnras,
  400, 299

\bibitem[\protect\citeauthoryear{{Ludlam}, {Cackett}, {G{\"u}ltekin} \& et
  al.}{{Ludlam} et~al.}{2015}]{2015MNRAS.447.2112L}
{Ludlam} R.~M.,  {Cackett} E.~M.,  {G{\"u}ltekin} K.,    et al. 2015, \mnras,
  447, 2112

\bibitem[\protect\citeauthoryear{{Lusso}, {Worseck}, {Hennawi}, {Prochaska},
  {Vignali}, {Stern} \& {O'Meara}}{{Lusso} et~al.}{2015}]{2015MNRAS.449.4204L}
{Lusso} E.,  {Worseck} G.,  {Hennawi} J.~F.,  {Prochaska} J.~X.,  {Vignali} C.,
   {Stern} J.,    {O'Meara} J.~M.,  2015, \mnras, 449, 4204

\bibitem[\protect\citeauthoryear{{Madau} \& {Haardt}}{{Madau} \&
  {Haardt}}{2015}]{2015ApJ...813L...8M}
{Madau} P.,  {Haardt} F.,  2015, \apjl, 813, L8

\bibitem[\protect\citeauthoryear{{Masters}, {Capak}, {Salvato}, {Civano},
  {Mobasher}, {Siana}, {Hasinger}, {Impey}, {Nagao}, {Trump}, {Ikeda}, {Elvis}
  \& {Scoville}}{{Masters} et~al.}{2012}]{2012ApJ...755..169M}
{Masters} D.,  {Capak} P.,  {Salvato} M.,  {Civano} F.,  {Mobasher} B.,
  {Siana} B.,  {Hasinger} G.,  {Impey} C.~D.,  {Nagao} T.,  {Trump} J.~R.,
  {Ikeda} H.,  {Elvis} M.,    {Scoville} N.,  2012, \apj, 755, 169

\bibitem[\protect\citeauthoryear{{Matsuda}, {Yamada}, {Hayashino} \& et.
  al.}{{Matsuda} et~al.}{2012}]{2012MNRAS.425..878M}
{Matsuda} Y.,  {Yamada} T.,  {Hayashino} T.,    et. al. 2012, \mnras, 425, 878

\bibitem[\protect\citeauthoryear{{McHardy}, {Gunn}, {Uttley} \& et
  al.}{{McHardy} et~al.}{2005}]{2005MNRAS.359.1469M}
{McHardy} I.~M.,  {Gunn} K.~F.,  {Uttley} P.,    et al. 2005, \mnras, 359, 1469

\bibitem[\protect\citeauthoryear{{Meiksin}}{{Meiksin}}{2005}]{2005MNRAS.356..596M}
{Meiksin} A.,  2005, \mnras, 356, 596

\bibitem[\protect\citeauthoryear{{Micheva}, {Iwata}, {Inoue}, {Matsuda},
  {Yamada} \& {Hayashino}}{{Micheva} et~al.}{2015}]{2015arXiv150903996M}
{Micheva} G.,  {Iwata} I.,  {Inoue} A.~K.,  {Matsuda} Y.,  {Yamada} T.,
  {Hayashino} T.,  2015, ArXiv e-prints

\bibitem[\protect\citeauthoryear{{Miniutti}, {Ponti}, {Greene} \& et
  al.}{{Miniutti} et~al.}{2009}]{2009MNRAS.394..443M}
{Miniutti} G.,  {Ponti} G.,  {Greene} J.~E.,    et al. 2009, \mnras, 394, 443

\bibitem[\protect\citeauthoryear{{Morokuma}, {Doi}, {Yasuda} \& et
  al.}{{Morokuma} et~al.}{2008}]{2008ApJ...676..163M}
{Morokuma} T.,  {Doi} M.,  {Yasuda} N.,    et al. 2008, \apj, 676, 163

\bibitem[\protect\citeauthoryear{{Nestor}, {Shapley}, {Kornei} \& et.
  al.}{{Nestor} et~al.}{2013}]{2013ApJ...765...47N}
{Nestor} D.~B.,  {Shapley} A.~E.,  {Kornei} K.~A.,    et. al. 2013, \apj, 765,
  47

\bibitem[\protect\citeauthoryear{{Nestor}, {Shapley}, {Steidel} \&
  {Siana}}{{Nestor} et~al.}{2011}]{2011ApJ...736...18N}
{Nestor} D.~B.,  {Shapley} A.~E.,  {Steidel} C.~C.,    {Siana} B.,  2011, \apj,
  736, 18

\bibitem[\protect\citeauthoryear{{Netzer}, {Kaspi}, {Behar} \& et al.}{{Netzer}
  et~al.}{2003}]{2003ApJ...599..933N}
{Netzer} H.,  {Kaspi} S.,  {Behar} E.,    et al. 2003, \apj, 599, 933

\bibitem[\protect\citeauthoryear{{Palanque-Delabrouille}, {Magneville},
  {Y{\`e}che} \& et al.}{{Palanque-Delabrouille}
  et~al.}{2013}]{2013A&A...551A..29P}
{Palanque-Delabrouille} N.,  {Magneville} C.,  {Y{\`e}che} C.,    et al. 2013,
  \aap, 551, A29

\bibitem[\protect\citeauthoryear{{Parker}, {Fabian}, {Matt} \& et al.}{{Parker}
  et~al.}{2015}]{2015MNRAS.447...72P}
{Parker} M.~L.,  {Fabian} A.~C.,  {Matt} G.,    et al. 2015, \mnras, 447, 72

\bibitem[\protect\citeauthoryear{{Pentericci}, {Fan}, {Rix} \& et.
  al.}{{Pentericci} et~al.}{2002}]{2002AJ....123.2151P}
{Pentericci} L.,  {Fan} X.,  {Rix} H.-W.,    et. al. 2002, \aj, 123, 2151

\bibitem[\protect\citeauthoryear{{Ponti}, {Papadakis}, {Bianchi} \& et
  al.}{{Ponti} et~al.}{2012}]{2012A&A...542A..83P}
{Ponti} G.,  {Papadakis} I.,  {Bianchi} S.,    et al. 2012, \aap, 542, A83

\bibitem[\protect\citeauthoryear{{Prochaska}, {Hennawi}, {Lee}, {Cantalupo},
  {Bovy}, {Djorgovski}, {Ellison}, {Lau}, {Martin}, {Myers}, {Rubin} \&
  {Simcoe}}{{Prochaska} et~al.}{2013}]{2013ApJ...776..136P}
{Prochaska} J.~X.,  {Hennawi} J.~F.,  {Lee} K.-G.,  {Cantalupo} S.,  {Bovy} J.,
   {Djorgovski} S.~G.,  {Ellison} S.~L.,  {Lau} M.~W.,  {Martin} C.~L.,
  {Myers} A.,  {Rubin} K.~H.~R.,    {Simcoe} R.~A.,  2013, \apj, 776, 136

\bibitem[\protect\citeauthoryear{{Razoumov} \& {Sommer-Larsen}}{{Razoumov} \&
  {Sommer-Larsen}}{2010}]{2010ApJ...710.1239R}
{Razoumov} A.~O.,  {Sommer-Larsen} J.,  2010, \apj, 710, 1239

\bibitem[\protect\citeauthoryear{{Saez}, {Lehmer}, {Bauer} \& et al.}{{Saez}
  et~al.}{2015}]{2015MNRAS.450.2615S}
{Saez} C.,  {Lehmer} B.~D.,  {Bauer} F.~E.,    et al. 2015, \mnras, 450, 2615

\bibitem[\protect\citeauthoryear{{Sarajedini}, {Koo}, {Klesman} \& et
  al.}{{Sarajedini} et~al.}{2011}]{2011ApJ...731...97S}
{Sarajedini} V.~L.,  {Koo} D.~C.,  {Klesman} A.~J.,    et al. 2011, \apj, 731,
  97

\bibitem[\protect\citeauthoryear{{Schaerer}}{{Schaerer}}{2003}]{2003A&A...397..527S}
{Schaerer} D.,  2003, \aap, 397, 527

\bibitem[\protect\citeauthoryear{{Schlafly} \& {Finkbeiner}}{{Schlafly} \&
  {Finkbeiner}}{2011}]{2011ApJ...737..103S}
{Schlafly} E.~F.,  {Finkbeiner} D.~P.,  2011, \apj, 737, 103

\bibitem[\protect\citeauthoryear{{Schulze}, {Wisotzki} \& {Husemann}}{{Schulze}
  et~al.}{2009}]{2009A&A...507..781S}
{Schulze} A.,  {Wisotzki} L.,    {Husemann} B.,  2009, \aap, 507, 781

\bibitem[\protect\citeauthoryear{{Scott}, {Kriss}, {Brotherton}, {Green},
  {Hutchings}, {Shull} \& {Zheng}}{{Scott} et~al.}{2004}]{2004ApJ...615..135S}
{Scott} J.~E.,  {Kriss} G.~A.,  {Brotherton} M.,  {Green} R.~F.,  {Hutchings}
  J.,  {Shull} J.~M.,    {Zheng} W.,  2004, \apj, 615, 135

\bibitem[\protect\citeauthoryear{{Shull}, {Moloney}, {Danforth} \& et.
  al.}{{Shull} et~al.}{2015}]{2015ApJ...811....3S}
{Shull} J.~M.,  {Moloney} J.,  {Danforth} C.~W.,    et. al. 2015, \apj, 811, 3

\bibitem[\protect\citeauthoryear{{Steidel}, {Erb}, {Shapley}, {Pettini},
  {Reddy}, {Bogosavljevi{\'c}}, {Rudie} \& {Rakic}}{{Steidel}
  et~al.}{2010}]{2010ApJ...717..289S}
{Steidel} C.~C.,  {Erb} D.~K.,  {Shapley} A.~E.,  {Pettini} M.,  {Reddy} N.,
  {Bogosavljevi{\'c}} M.,  {Rudie} G.~C.,    {Rakic} O.,  2010, \apj, 717, 289

\bibitem[\protect\citeauthoryear{{Telfer}, {Zheng}, {Kriss} \&
  {Davidsen}}{{Telfer} et~al.}{2002}]{2002ApJ...565..773T}
{Telfer} R.~C.,  {Zheng} W.,  {Kriss} G.~A.,    {Davidsen} A.~F.,  2002, \apj,
  565, 773

\bibitem[\protect\citeauthoryear{{Thornton}, {Barth}, {Ho} \& et
  al.}{{Thornton} et~al.}{2008}]{2008ApJ...686..892T}
{Thornton} C.~E.,  {Barth} A.~J.,  {Ho} L.~C.,    et al. 2008, \apj, 686, 892

\bibitem[\protect\citeauthoryear{{Trevese}, {Boutsia}, {Vagnetti} \& et
  al.}{{Trevese} et~al.}{2008}]{2008A&A...488...73T}
{Trevese} D.,  {Boutsia} K.,  {Vagnetti} F.,    et al. 2008, \aap, 488, 73

\bibitem[\protect\citeauthoryear{{Tr{\`e}vese} \& {Vagnetti}}{{Tr{\`e}vese} \&
  {Vagnetti}}{2002}]{2002ApJ...564..624T}
{Tr{\`e}vese} D.,  {Vagnetti} F.,  2002, \apj, 564, 624

\bibitem[\protect\citeauthoryear{{Ulrich}, {Maraschi} \& {Urry}}{{Ulrich}
  et~al.}{1997}]{1997ARA&A..35..445U}
{Ulrich} M.-H.,  {Maraschi} L.,    {Urry} C.~M.,  1997, \araa, 35, 445

\bibitem[\protect\citeauthoryear{{Vanden Berk}, {Richards}, {Bauer} \& et.
  al.}{{Vanden Berk} et~al.}{2001}]{2001AJ....122..549V}
{Vanden Berk} D.~E.,  {Richards} G.~T.,  {Bauer} A.,    et. al. 2001, \aj, 122,
  549

\bibitem[\protect\citeauthoryear{{Vanzella}, {de Barros}, {Castellano} \& et
  al.}{{Vanzella} et~al.}{2015}]{2015A&A...576A.116V}
{Vanzella} E.,  {de Barros} S.,  {Castellano} M.,    et al. 2015, \aap, 576,
  A116

\bibitem[\protect\citeauthoryear{{Webb}, {Yamada}, {Huang} \& et. al.}{{Webb}
  et~al.}{2009}]{2009ApJ...692.1561W}
{Webb} T.~M.~A.,  {Yamada} T.,  {Huang} J.-S.,    et. al. 2009, \apj, 692, 1561

\bibitem[\protect\citeauthoryear{{Wilman}, {Gerssen}, {Bower}, {Morris},
  {Bacon}, {de Zeeuw} \& {Davies}}{{Wilman} et~al.}{2005}]{2005Natur.436..227W}
{Wilman} R.~J.,  {Gerssen} J.,  {Bower} R.~G.,  {Morris} S.~L.,  {Bacon} R.,
  {de Zeeuw} P.~T.,    {Davies} R.~L.,  2005, \nat, 436, 227

\bibitem[\protect\citeauthoryear{{Yamada}, {Nakamura}, {Matsuda} \& et.
  al.}{{Yamada} et~al.}{2012}]{2012AJ....143...79Y}
{Yamada} T.,  {Nakamura} Y.,  {Matsuda} Y.,    et. al. 2012, \aj, 143, 79

\bibitem[\protect\citeauthoryear{{Zheng}, {Kriss}, {Telfer}, {Grimes} \&
  {Davidsen}}{{Zheng} et~al.}{1997}]{1997ApJ...475..469Z}
{Zheng} W.,  {Kriss} G.~A.,  {Telfer} R.~C.,  {Grimes} J.~P.,    {Davidsen}
  A.~F.,  1997, \apj, 475, 469

\end{thebibliography}

\protect\label{lastpage}
\end{document}